\documentclass[amssymb,aps,pra,superscriptaddress, reprint, floatfix]{revtex4-1}

\usepackage{graphicx}
\usepackage{float}
\usepackage{longtable}
\usepackage{mathtools}
\usepackage{tikz}
\usepackage{siunitx}
\usepackage{xcolor}
\usepackage[normalem]{ulem}
\usepackage{xr}
\begin{document}

\title{Morphology and Motility of Cells on Soft Substrates}
\author{Andriy Goychuk}
\affiliation{Arnold Sommerfeld Center for Theoretical Physics and Center for NanoScience, Department of Physics, Ludwig-Maximilians-Universit\"at M\"unchen, Theresienstr. 37, D-80333 Munich, Germany}
\author{David B. Br\"uckner}
\affiliation{Arnold Sommerfeld Center for Theoretical Physics and Center for NanoScience, Department of Physics, Ludwig-Maximilians-Universit\"at M\"unchen, Theresienstr. 37, D-80333 Munich, Germany}
\author{Andrew W. Holle}
\affiliation{Department of Cellular Biophysics, Max-Planck-Institute for Medical Research, D-69120 Heidelberg, Germany}
\author{Joachim P. Spatz}
\affiliation{Department of Cellular Biophysics, Max-Planck-Institute for Medical Research, D-69120 Heidelberg, Germany}
\affiliation{Department of Biophysical Chemistry, University of Heidelberg, D-69120 Heidelberg, Germany}
\author{Chase P. Broedersz}
\affiliation{Arnold Sommerfeld Center for Theoretical Physics and Center for NanoScience, Department of Physics, Ludwig-Maximilians-Universit\"at M\"unchen, Theresienstr. 37, D-80333 Munich, Germany}
\author{Erwin Frey}
\affiliation{Arnold Sommerfeld Center for Theoretical Physics and Center for NanoScience, Department of Physics, Ludwig-Maximilians-Universit\"at M\"unchen, Theresienstr. 37, D-80333 Munich, Germany}

\begin{abstract}
Recent experiments suggest that the interplay between cells and the mechanics of their substrate gives rise to a diversity of morphological and migrational behaviors.
Here, we develop a Cellular Potts Model of polarizing cells on a visco-elastic substrate.
We compare our model with experiments on endothelial cells plated on polyacrylamide hydrogels to constrain model parameters and test predictions.
Our analysis reveals that morphology and migratory behavior are determined by an intricate interplay between cellular polarization and substrate strain gradients generated by traction forces exerted by cells (self-haptotaxis). 
\end{abstract}

\maketitle

\let\oldaddcontentsline\addcontentsline
\renewcommand{\addcontentsline}[3]{}

Cell migration is a highly complex process determined by internal chemo-mechanical processes and the interaction of the cell with its environment~\cite{Charras:2014ht, DePascalis:2017dy, Silberzan:2017ww, vanHelvert:2018}.
Indeed, cells respond to the mechanical properties of the substrate to which they adhere~\cite{Lo:2000, Isenberg:2009, Zemel:2010a, Zemel:2010b, Vincent:2013ut, Hartman:2016, Sunyer:2016, Hadden:2017, Lachowski:2017uw}.
Interestingly, with increasing substrate rigidity, different cell types show qualitatively distinct migratory behavior. 
For example, glioma cells~\cite{Ulrich:2009}, glioblastoma cells~\cite{Grundy:2016}, and human adipose-derived stem cells~\cite{Hadden:2017} plated on polyacrylamide (PA) hydrogels, as well as fish keratocytes on PA and polydimethylsiloxane (PDMS) hydrogels~\cite{Riaz:2016}, move faster and more persistently with increasing elastic modulus. 
In contrast, rat fibroblasts plated on polyethylene glycol-based (PEG) hydrogels~\cite{Missirlis:2014}, as well as 3T3 fibroblasts on PA hydrogels~\cite{Pelham:1997}, show the opposite behavior and slow down, while still increasing their persistence of migration on stiffer substrates.
What then are the physical principles that lead to such diverse cell behaviors?

Substrates like PA and PEG hydrogels are widely regarded as almost ideally elastic materials ~\cite{Calvet:2004, Raeber:2005}. 
In general, however, substrate viscosity may also affect cell migration. 
For example, correlations in the movement of epithelial sheets have been shown to increase with substrate viscosity~\cite{Murrell:2011}, and a recent computational study has demonstrated the relevance of viscous substrate remodelling for cell spreading~\cite{Chaudhuri:2015}.
These studies suggest an intricate interplay between cell migration and both the elastic and viscous properties of the environment.
It remains to be resolved, however, whether and how these cell-substrate interactions can reconcile the apparently contradictory migratory responses of various cell types on different substrates. 

Previous computational approaches, including phase field models~\cite{Ziebert:2012, Shao:2012, Ziebert:2013, Camley:2013, Camley:2014a, Camley:2014b, Lober:2015}, cellular Potts models (CPM)~\cite{Graner:1992, Albert:2014, vanOers:2014, Thueroff:2017, Segerer:2015, Albert:2016, Rens:2017dn}, particle-based models~\cite{Viscek:1995, Zaman:2005up, Sepulveda:2013, Barnhart:2011, Garcia:2015, Chepizhko:2016, Novikova:2017, Schnyder:2017ws, Dietrich:2018}, and various continuum models~\cite{Oster:1983, Murray:1983, Besser:2010fe, Zemel:2010a, Zemel:2010b, Harland:2011, Friedrich:2012jv, EloseguiArtola:2014hg, EloseguiArtola:2016, Sunyer:2016, Bennett:2018cs}, have led to important advances in understanding cell traction force generation and cell migration.
In particular, these studies have helped to rationalize the coupling between single-cell motion and substrate deformation~\cite{Oster:1983, Murray:1983, Ziebert:2013, vanOers:2014, EloseguiArtola:2014hg, EloseguiArtola:2016, Sunyer:2016, Rens:2017dn, Dietrich:2018}. 
However, these models neglect spatial coupling of substrate deformations~\cite{Ziebert:2013}, cannot capture cell shape~\cite{Oster:1983, Murray:1983, Sunyer:2016, Dietrich:2018}, do not include a cell polarization mechanism~\cite{Oster:1983, Murray:1983, vanOers:2014, Sunyer:2016, Rens:2017dn}, and mostly exclude persistent cell migration.

\begin{figure}[b]
	\includegraphics{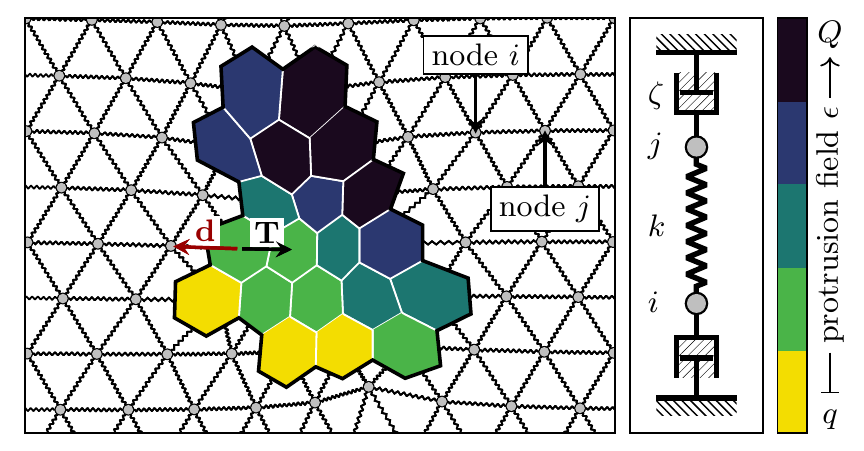}
	\caption{Sketch of the computational model.
		The substrate is represented by nodes $i$ at positions $\mathbf{x}_i$, each connected to six nearest neighbors $j \,{\in}\, \mathcal{N}_i$ by loaded springs. 
		A cell $\cal{C}$ is comprised of a set of hexagons with respective areas $a(\mathbf{x}_i,t)$ and local protrusion energies $\epsilon (\mathbf{x}_i,t) \,{\in}\, [q,Q]$ (color scale). 
		As the cell exerts traction forces $\mathbf{T}$ on the nodes, it compresses the substrate beneath, while stretching the surrounding substrate.
		The cell protrudes or retracts over an effective distance $\vert\mathbf{d}\vert$ in the direction $\pm\mathbf{d}$, where $\mathbf{d} \,{=}\, \mathbf{x}_j {-} \mathbf{x}_i$.
		}
	\label{fig::illustration}
\end{figure}

Here, we study the morphology and migratory behavior of actively polarizing cells on visco-elastic substrates of varying elastic stiffness and different degrees of viscous friction. 
To this end, we develop a CPM of actively polarizing motile cells~\cite{Thueroff:2017, Segerer:2015} that mechanically interact with a simple visco-elastic substrate [Fig.~\ref{fig::illustration}], using experimental measurements on human umbilical vein endothelial cells (HUVECs) plated on PA gels to constrain the model parameters.
Our combined experimental and theoretical investigations suggest that a cell's response to the physical properties of the substrate can be understood in a relatively simple way, without explicitly taking into account additional effects like stiffness-dependent adhesions. 
Within our picture, cells generate substrate strain gradients, which guide shape changes and cell migration (self-haptotaxis).
The interaction with the substrate can in turn interfere with, and even override, internal feedback mechanisms that would under normal circumstances lead to cell polarization. 

\begin{figure}[b]
	\includegraphics{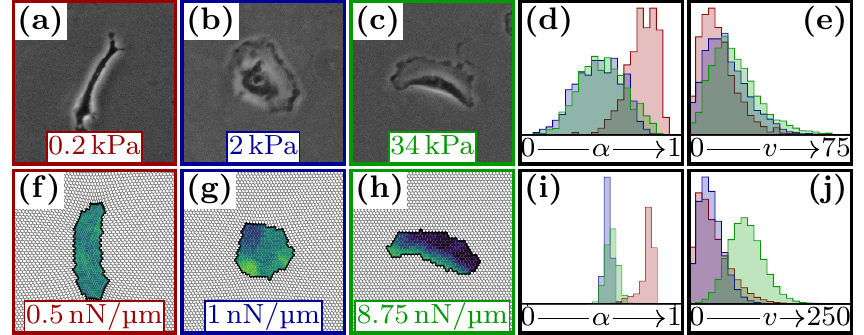}
	\caption{
		{Migratory cell states.} 
		Depending on substrate stiffness, HUVECs show distinct migratory cell states: (a) \textit{elongation}, (b) \textit{rounding},  and (c) \textit{running}, as quantified by the histograms for the cell extension $\alpha \,{=}\, 1\,{-}\,4 \pi A/P^2$ (d), and local cell speed $v$ (e) in [\si[per-mode=symbol]{\nano\meter\per\second}]).
		(f)-(j) Analogous results from the computational model for viscous friction $\zeta = \SI[per-mode=symbol]{17.5}{\second\nano\newton\per\micro\meter}$.
		}
	\label{fig::states}
\end{figure}

\textit{Experimental observations.} We started our analysis by experimentally investigating HUVECs plated on PA gels.
Depending on the substrate stiffness, we observed three distinct migratory cell behaviors [Fig.~\ref{fig::states}(a)-(e)].
At low stiffness, cells are elongated and localized (\textit{elongation}): Even though they locally move at some slow speed $v$ in random directions, they remain localized within a certain substrate area and do not show persistent motion.
As substrate stiffness is increased, cells first round up and increase their local speed, but remain localized (\textit{rounding}). 
Only when the substrate stiffness is increased above some threshold value, do cells begin to show persistent cell migration (\textit{running}), which can be described as a persistent random walk with ballistic motion on short timescales and diffusive motion on long timescales.

\textit {Generalized CPM.} 
To rationalize these diverse cell behaviors we build on and extend a recently introduced generalization of the CPM~\cite{Thueroff:2017, Segerer:2015}, which includes the following basic features of cellular dynamics: 
Elasticity of the cell membrane and cortex, dynamic cell polarization through a chemo-mechanical feedback mechanism, and force generation driven by the interplay between actin polymerization and contraction of acto-myosin networks. 
As described below, we add as a new feature the visco-elastic coupling of cell and substrate deformations.

We consider a cell as a connected set $\mathcal{C}$ of occupied grid sites (hexagons) $i$, with positions $\mathbf{x}_i (t)$ and areas $a(\mathbf{x}_i,t)$ that change with time $t$ [Fig.~\ref{fig::illustration}]. 
Cell motion and cell shape changes are implemented as  
elementary protrusion and retraction events, corresponding to an increase and decrease, respectively, in the number of occupied grid sites. 
Moreover, through coupling with a visco-elastic substrate, the individual grid areas may change dynamically.
The dynamics of each cell is determined by a Monte Carlo update scheme with the `statistical' weight being given by a Boltzmann factor with a Hamiltonian $\mathcal{H} \,{=}\, \mathcal{H}_P {+} \mathcal{H}_M$, which describes the balance between a cell's tendency to protrude and migrate and the constraints imposed by membrane elasticity.

As in the original CPM~\cite{Graner:1992}, deformations of a cell's membrane and cortex are assumed to be constrained by the elastic energy $\mathcal{H}_{M} \,{=}\, \kappa_{A} A(t)^{2} {+} \kappa_{P} P(t)^{2}$, with $\kappa_A$ and $\kappa_P$ denoting the stiffnesses corresponding to the area $A(t)$ and perimeter $P(t)$ of the cell, respectively.
The ensuing contractile forces are counteracted by outwardly directed forces generated by cytoskeletal structures anchored to the substrate at focal adhesion sites~\cite{Pollard:2003, Mogilner:2009}. 
In our model, the local energetic contribution from this cellular activity is described by $\mathcal{H}_P \,{=}\, {-} \sum_{i \in \mathcal{C}} \epsilon(\mathbf{x}_i,t)$, with the scalar \textit{protrusion} field $\epsilon(\mathbf{x}_i,t) \,{\in}\, [q, Q]$~\cite{Thueroff:2017, Segerer:2015}. 
The protrusion field is dynamic, reflecting the response of cytoskeletal structures to external mechanical stimuli through feedback mechanisms involving regulatory cytoskeletal proteins~\cite{Maree:2006, Maree:2012}.
In the generalized CPM these complex biochemical processes are accounted for in a simplified way by regulatory factors that reinforce the protrusion field in a positive feedback loop that can lead to spontaneous cell polarization~\cite{Segerer:2015, Thueroff:2017}; for details see the Supplemental Material (S.M.)~\cite{supplement}.

\textit {Cell-substrate coupling.} 
How can one account for substrate deformations and their coupling to cell deformation in a CPM?
As the cell's cytoskeleton is anchored to the substrate via focal adhesion sites while the cell is exerting force on the substrate, we will assume that each hexagon area $a(\mathbf{x}_i,t)$ deforms in an affine way with the substrate. 
In the continuum limit, this implies that the protrusion energy density is given by $\epsilon(\mathbf{x},t)/a(\mathbf{x},t)$, and the total protrusion energy can thus be written as an integral over the cell area $A$:
\begin{equation}
	\mathcal{H}_{P} 
	= - \int_{A} \! \frac{\mathrm{d}^2\mathbf{x}}{a_0} \; \epsilon(\mathbf{x},t) \,\sigma(\mathbf{x},t) \, ,
\label{eq::hamiltonian_polarization}
\end{equation}
where $\sigma(\mathbf{x},t) \,{=}\, a_0/a(\mathbf{x},t)$ represents the local compression or dilatation with respect to the area of an undeformed hexagon $a_0$. 
Hence $\mathcal{H}_{P}$ favors high protrusion energy density $\epsilon(\mathbf{x},t) \, \sigma(\mathbf{x},t)$.
 
When a cell attempts to protrude/retract in the direction $\mathbf{\pm d}$ [Fig.~\ref{fig::illustration}], the forces $\mathbf{F}$ facilitating this effort are balanced (on the scale of the grid sites) by traction forces $\mathbf{T}$. 
For instance, during a protrusion, the actin cytoskeleton exerts a pushing force $\mathbf{F}_P(\mathbf{x}_i,t)$ over the distance $|\mathbf{d}(\mathbf{x}_i,t)|$, which is determined by a change in polarization energy accounted for by Eq.~\eqref{eq::hamiltonian_polarization}. 
This pushing force is transmitted to the substrate by focal adhesions and balanced locally by a traction force $\mathbf{T}_P \,{=}\, {-}\mathbf{F}_P$, which is directed towards the cell interior:
\begin{equation}
	\mathbf{T}_P(\mathbf{x}_i,t) = - \frac{\vert\Delta \mathcal{H}_P(\mathbf{x}_i,t)\vert}{\vert\mathbf{d}(\mathbf{x}_i,t)\vert^2} 
	    \, \mathbf{d}(\mathbf{x}_i,t) \, .
\end{equation}
Similarly, the change in cell morphological energy associated with a protrusion or retraction can be related to an effective contractile force on the cell membrane:
\begin{equation}
	\mathbf{F}_M(\mathbf{x}_i,t) = -\frac{|\Delta \mathcal{H}_M (\mathbf{x}_i,t)|}{\vert\mathbf{d}(\mathbf{x}_i,t)\vert^2} \, \mathbf{d}(\mathbf{x}_i,t) \, .
\end{equation}
We assume that the cytoskeleton facilitates this contractility by transmitting forces instantaneously throughout the cell~\cite{Wang:2009iv}.
Then, the contractile force $\mathbf{F}_M$ is distributed homogeneously over all hexagons $j\in\mathcal{C}$ occupied by the cell and balanced by traction forces $\mathbf{T}_M(\mathbf{x}_j) \,{=}\, {-}\mathbf{F}_M(\mathbf{x}_i)/\vert\mathcal{C}\vert$.

In the course of spreading and migration, the cell exerts the traction forces $\mathbf{T}(\mathbf{x}_i) \,{=}\, \mathbf{T}_P(\mathbf{x}_i) \,{+}\, \mathbf{T}_M(\mathbf{x}_i)$ on the nodes $\mathbf{x}_i$ of the substrate, which is described as a discrete network of beads [Fig.~\ref{fig::illustration}] subject to viscous damping with viscous friction coefficient $\zeta$~\cite{Yucht:2013iw}, and connected by loaded springs with spring coefficient $k$; for details see S.M.~\cite{supplement}.
Force balance then determines the overdamped dynamics for each node: $\zeta \, \dot{\mathbf{x}}_i (t) \,{=}\,  \mathbf{T}(\mathbf{x}_i) \,{+}\, k \sum_{j\in \mathcal{N}_i} \, (\mathbf{x}_j \,{-}\, \mathbf{x}_i)$.

\textit{Parameter estimation.}
To compare the experimental results with our computational model, we chose the model parameters to ensure physiological values for the cell speed $v$, spreading area $A$ and traction forces on the substrate~\cite{supplement}.
We determined the range of studied spring coefficients $k$ to match the elastic properties of the substrate.
Specifically, a spring coefficient of $k \,{=}\, \SI[per-mode = symbol]{0.5}{\nano\newton\per\micro\meter}$ corresponds to a substrate modulus of $E \,{\approx}\, \SI{0.6}{\kilo\pascal}$~\cite{supplement}.
For the choice of the friction coefficient $\zeta$ we distinguish between two representative cases, depending on the relative timescales for relaxation of the visco-elastic network ($\tau^{}_R \,{=}\, \zeta / k$) and cell migration ($\tau_\mathcal{C}$). 
Since a lower bound for $\tau_\mathcal{C}$ is given by the inverse update rate of internal cell polarization, $\tau_\mathcal{C} \,{\geq}\, 1/g_\epsilon$, we expect viscous friction effects to become significant at $\zeta^\star \,{\approx}\, \SI[per-mode=symbol] {35}{\second \nano\newton\per\micro\meter}$.
This motivates our choice of the representative values $\zeta \,{=}\, \SI[per-mode=symbol]{121}{\second\nano\newton\per\micro\meter}$ and  $\zeta \,{=}\, \SI[per-mode=symbol]{17.5}{\second\nano\newton\per\micro\meter}$ for what we call high and low substrate viscosity, respectively, in the following. 
A table of the parameter values of the CPM is given in the S.M.~\cite{supplement}.

\begin{figure}[t]
	\includegraphics{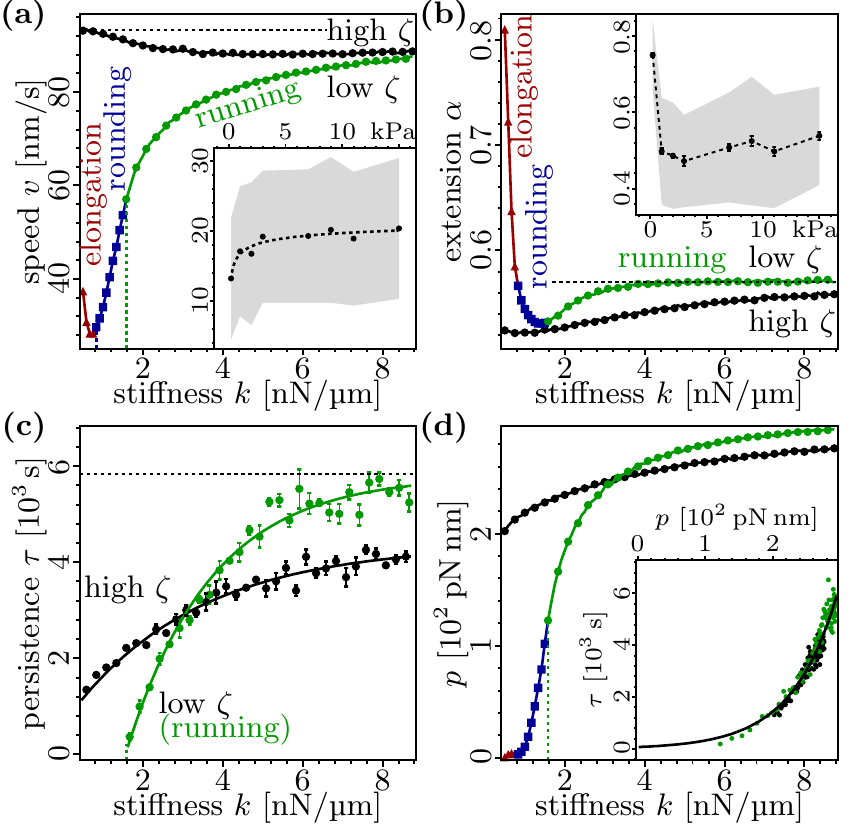}
	\caption{
		Characterization of cell migration and morphology. For a reference, horizontal dashed lines indicate the cell behavior for rigid substrates. All lines are guides to the eye.
		(a) For $\zeta \,{=}\, \SI[per-mode=symbol]{121}{\second\nano\newton\per\micro\meter}$ (high $\zeta$, filled black circles), the local cell speed $v$ decreases with increasing substrate stiffness $k$. In contrast, and in accordance with experimental results for HUVEC cells on PA gels (\textit{Inset}), $v$ increases with stiffness for $\zeta \,{=}\, \SI[per-mode=symbol]{17.5}{\second\nano\newton\per\micro\meter}$ (low $\zeta$). There are three distinct migratory cell states: 	\textit{elongation} (red triangles), \textit{rounding} (blue squares), and \textit{running} (green circles).
		(b) For low $\zeta$, both experiment (\textit{Inset}) and the theoretical model show biphasic behavior in the cell extension $\alpha$ with pronounced elongation for low $k$.
		(c) The persistence time $\tau$ of directed cell migration, as obtained from fitting a persistent random walk to the cell trajectories, shows a positive correlation with $k$.	
		(d) Cell polarization $p$ as a function of substrate stiffness for low and high substrate viscosity. \textit{Inset:} Correlation plot of  persistence time $\tau$ versus cell polarization $p$.
	}
	\label{fig::stiffness_study}
\end{figure}

\textit{Low viscous friction.}
Because PA gels are dominated by their elastic and not their viscous properties~\cite{Calvet:2004}, we used a low value $\zeta \,{=}\, \SI[per-mode=symbol]{17.5}{\second\nano\newton\per\micro\meter}$, below the threshold value $\zeta^\star$, and find the same phenomenology as in our experiments. 
Not only does our model capture the distinct morphologies at different substrate stiffness [Fig.~\ref{fig::states}~(f)-(j)], it also accounts for the onset of motility, i.e. the transition towards the running state beyond a threshold in substrate stiffness [Fig.~\ref{fig::stiffness_study}(a)].
Indeed, our simulations are consistent with experiments which show that the cell's speed increases with substrate  stiffness [inset of Fig.~\ref{fig::stiffness_study}(a)].
Previous experiments~\cite{Riaz:2016} show a monotonic increase in cell elongation for high substrate rigidity (${>}\SI[per-mode=symbol]{1.5}{\kilo\pascal}$). 
We observe the same monotonic trend in our experiments with HUVECs plated on PA gels of comparable rigidity (${>}\SI[per-mode=symbol]{1}{\kilo\pascal}$).
Interestingly, extending these measurements to low substrate rigidity (less than $\SI[per-mode=symbol]{1}{\kilo\pascal}$), we observe pronounced cell elongation [inset of Fig.~\ref{fig::stiffness_study}(b)].
This biphasic behavior is fully in accordance with our computer simulations without any further adjustment of parameters [Fig.~\ref{fig::stiffness_study}(b)].
Moreover, the computational model predicts that the persistence time $\tau$, as determined from fitting a persistent random walk to the cell trajectories, increases with substrate stiffness [Fig.~\ref{fig::stiffness_study}(c)], in full agreement with previous experimental results~\cite{Riaz:2016, Ulrich:2009, Grundy:2016}.

\textit{High viscous friction.}
We then looked at the effects of substrate viscosity on the migratory behavior of cells.
On raising the viscous friction coefficient $\zeta$ above the threshold value $\zeta^\star$, we find a considerable change in phenomenology [Fig.~\ref{fig::stiffness_study}(a)-(c)]:
Cells now only exhibit \textit{running} states, with cell speed decreasing, and both persistence time and elongation monotonically increasing with substrate stiffness. 
Qualitatively, these trends are similar to measurements of fibroblast motility on polyethylene glycol-based hydrogels~\cite{Missirlis:2014} and on PA gels~\cite{Pelham:1997}.

\textit{Cell phenotypes and substrate properties.}
How can one rationalize the dependence of the observed cellular phenotypes (morphology and motility) on substrate stiffness and viscosity [Fig.~\ref{fig::stiffness_study}] in terms of the interplay between substrate dynamics and cell polarization? 
Since by construction of the generalized CPM, a cell has highest probability to migrate in the direction of maximal protrusion energy density $\sigma \, \epsilon$, we analyzed correlations between this quantity and the persistence time.
We define the strength of cell polarization as
$p \,{=}\, \frac{1}{\pi} 
	\int_0^\pi\mathrm{d}\theta \, \cos \theta \,
	\epsilon_\text{b} (\theta, \theta_\epsilon) \,
	\sigma_\text{b} (\theta, \theta_\epsilon)
$
with $\theta$ being the angle relative to the average polarization axis $\theta_\epsilon$, and $\epsilon _\text{b} (\theta, \theta_\epsilon)$ and $\sigma_\text{b} (\theta, \theta_\epsilon)$ the protrusion energy and substrate density for hexagons along the interior boundary of a cell, respectively.

For high $\zeta$, a cell's persistence time $\tau$ remains finite even for very soft substrates [Fig.~\ref{fig::stiffness_study}(c)].
In contrast, for low $\zeta$, there is a threshold value $k^\star \,{\approx}\,  \SI[per-mode=symbol]{1.58}{\nano\newton\per\micro\meter}$ below which cells lose their persistence ($\tau \,{=}\, 0$) and become self-trapped. 
In this state, cells still show a finite average polarization [Fig.~\ref{fig::stiffness_study}(d)], but repolarize frequently, indicating that a threshold polarization strength is needed to sustain persistent cell migration against the substrate strain that tends to pull the cell back. 
Interestingly, we find that, regardless of the substrate properties, there is a universal increase of a cell's persistence time $\tau$ with cell polarization $p$, identifying it as the main determinant of the migratory persistence [inset of Fig.~\ref{fig::stiffness_study}(d)].

\begin{figure}[t]
\includegraphics{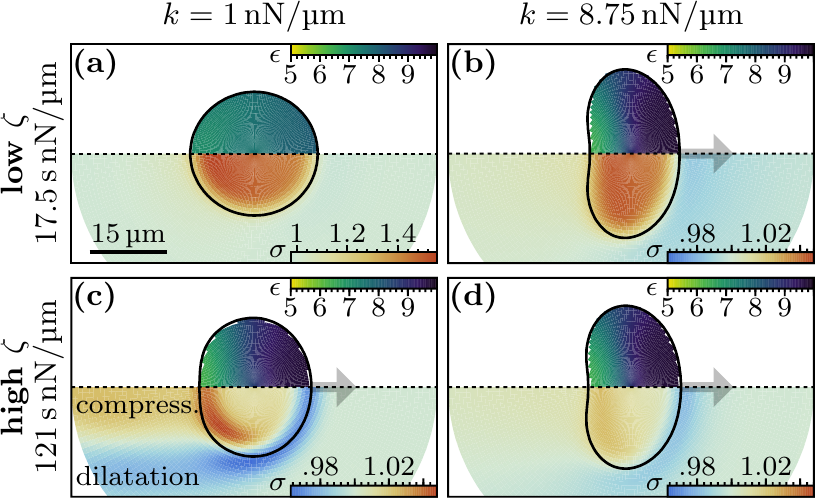}
	\caption{Cell polarization and substrate deformation. 
	Mirrored halves of a cell, with the local protrusion energy \textit{per hexagon} $\epsilon(x,y)$ [\SI[per-mode=symbol]{e2}{\pico\newton\nano\meter}] shown in the top halves and the substrate density (number of hexagons per unit area) $\sigma(x,-y)$ shown in the bottom halves. 
	These quantities are obtained by collapsing the data for many cells in their center of mass frame with the polarization axis oriented along the $x$-axis ($\theta_\epsilon \,{=}\, 0$).
	Note the differences in the scales of the substrate density $\sigma$.}
	\label{fig::states_simulation}
\end{figure}

Finally, we would like to illustrate how the interplay between cell polarization and substrate deformation leads to the different migratory states of a cell [Fig.~\ref{fig::states_simulation}]. 
Our simulations show that for low $\zeta$ and low substrate stiffness, cellular protrusion forces induce a strong compression of the substrate beneath the cell [Fig.~\ref{fig::states_simulation}(a)].
For a cell to move, it needs to protrude on one side and retract on the opposite side.
However, because substrate density is strongly increased below the cell, all retractions are energetically penalized.
Even in the event that a cell should manage to move, it would be energetically advantageous to simply move back to its previous position due to the local strain gradient (self-haptotaxis).
Due to the feedback between internal cell polarization and cell protrusions or retractions, inhibiting retractions effectively hampers cell polarization.
As a consequence, a cell stops performing a persistent random walk and becomes self-trapped on substrates with low stiffness. 
Moreover, as the cell is only transiently polarized, the bias in the individual protrusion and retraction rates is small, leading to a broad velocity distribution with respect to the (transient) axis of polarization and thereby to a low local cell speed. 
Conversely, for high values of substrate stiffness, substrate deformations are small and cells can polarize strongly [Fig.~\ref{fig::states_simulation}b]. 
Due to this strong polarization, cells migrate persistently [Fig.~\ref{fig::stiffness_study}(c)], and also at relatively high speeds [Fig.~\ref{fig::stiffness_study}(a)], as cell velocities are narrowly distributed with respect to the cell's polarization axis.

For high $\zeta$, the response of the substrate is slow compared to the intracellular dynamics.
Thus, to a first approximation, a cell behaves as if it were migrating on a completely rigid substrate, and hence can easily polarize even for low substrate stiffnesses [Fig.~\ref{fig::states_simulation}(c)].
The slow response of the substrate to cellular forces leads to a trail of increased  substrate density $\sigma$ behind the cell, and to a decrease in substrate density at the sides of the cell.
This leads to a lensing effect, which decreases the probability that the cell will deviate from a straight path.
This also explains why cell speed is enhanced at low substrate stiffness.
Moreover, the particular substrate density profile effectively reduces the cell polarization strength $p$ and with it the persistence time.
With increasing substrate stiffness all of these effects are attenuated as substrate deformations become smaller. As a consequence, cell speed decreases and persistence time increases, asymptotically approaching the corresponding values for low viscous friction of the substrate. 

\textit {Conclusion.} 
Though we cannot exclude gene regulation as a possible cause for distinct cellular responses to substrate stiffness and viscosity, our study shows that variability in cell behaviors can also be explained simply in terms of the physical properties of the substrate and its interplay with cell polarization.
This has potentially far-reaching consequences, as the mechanics of the physiological environment of cells varies depending on the tissue they are embedded in -- and this not only determines cell migration~\cite{vanHelvert:2018} but also stem cell differentiation and fate~\cite{Engler:2006ga, Akhmanova:2015}. 
Based on our results, one may speculate that the typical shape of cells (e.g. elongated 'neurons' at low stiffnesses, round 'adipocytes' at intermediate stiffnesses, 'keratocytes' at high stiffnesses) is not only pre-determined by gene regulation, but strongly affected by mechanical cross-talk with the extracellular matrix.

\begin{acknowledgments}
E.F. and C.B. acknowledge support by the German Excellence Initiative via the program `NanoSystems Initiative Munich' (NIM) and by the Deutsche Forschungsgemeinschaft (DFG) via Collaborative Research Center (SFB) 1032 (projects B02 and B12). 
A.G. and D.B.B. are supported by a DFG fellowship through the Graduate School of Quantitative Biosciences Munich (QBM). 
J.P.S. is the Weston Visiting Professor at the Weizmann Institute of Science and member of the cluster of excellence CellNetworks at Heidelberg University. 
J.P.S. and A.H. acknowledge support from the Max Planck Society. 
Parts of this work was performed at the Aspen Center for Physics, which is supported by National Science Foundation grant PHY-1607611.
\end{acknowledgments}

\cleardoublepage

\setcounter{page}{1}
\setcounter{figure}{0}
\renewcommand{\thefigure}{S\arabic{figure}}
\setcounter{equation}{0}
\renewcommand{\theequation}{S\arabic{equation}}
\renewcommand{\thetable}{S.\Roman{table}}
\renewcommand{\thesection}{S.\Roman{section}}

\onecolumngrid
\section*{Morphology and Motility of Cells on Soft Substrates -- Supporting Information}
\tableofcontents
\thispagestyle{empty}

\clearpage

\twocolumngrid
\appendix

\let\addcontentsline\oldaddcontentsline

\section{Numerical methods}

In the following sections, we describe the numerical methods employed in this paper and the detailed implementation of the model.
We first provide mathematical definitions for the substrate and the cell, recapitulate the Cellular Potts model and introduce our proposed extension to take substrate strains into account.
Furthermore, we give give concise definitions for our observables and an overview over all model parameters and their values.

\subsection{Mathematical description of the substrate}
\label{sec::substrate_description}

\begin{figure}[t]
	\includegraphics{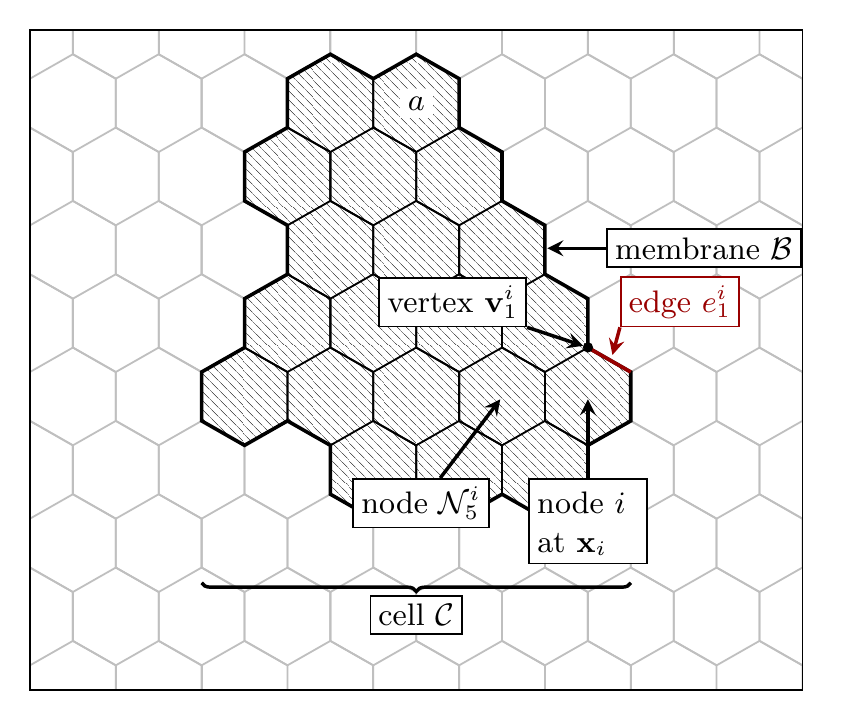}
	\caption{\textbf{Sketch of the cell and substrate morphology.} The substrate consists of hexagons with indices $i$ at positions $\mathbf{x}_i$.
	The vertices of the hexagons $\mathbf{v}_i(k)$ are obtained by interpolation, and the areas $a(\mathbf{x}_i,t)$ by the shoelace formula.
	The cell bulk is given by a morphologically connected set of hexagons $i \in\mathcal{C}$.
	The cell membrane $\mathcal{B}$ is the set of membrane segments $e_i(k)$ lining the border of the cell bulk.
	}
	\label{fig::morphology}
\end{figure}

The substrate (or \textit{grid}) is represented by a space-filling triangular lattice with time-dependent lattice vectors (or nodes) $\left\{\mathbf{x}_i(t)\right\}_{i=1,\dots,N}$.
This results in a hexagonal tesselation [Fig.~\ref{fig::morphology}] of the substrate consisting of $N$ hexagons with indices $i\in(1\dots N)$.
Each hexagonal tile $i$ is surrounded by six nearest neighbors that define the neighborhood $\mathcal{N}^i$:
\begin{equation}
	\mathcal{N}^{i} = \Bigl\{ j \, \bigl| \, \mathbf{x}_j(0) \text{ is nearest neighbor of } \mathbf{x}_i(0) \Bigr\}
\end{equation}
Strains in the substrate are modeled by deviations of the lattice vectors $\mathbf{x}_i(t)$ from the unstrained state.
In this unstrained state (at $t=0$ or for an infinitely stiff and undeformable substrate $k \,{\rightarrow}\, \infty$), nearest neighbors have a fixed distance from each other:
\begin{equation}
	|\mathbf{x}_j(0)-\mathbf{x}_i(0)|= d_0 \, \iff \, j \in {\mathcal{N}^{i}} \, .
\end{equation}
Furthermore, we impose a (clockwise) cyclic order on the set $\mathcal{N}^{i}$ with respect to the center tile $i$:
\begin{figure}[H]
	\centering
	\includegraphics{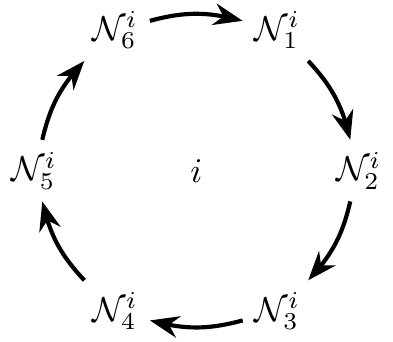}
\end{figure}

Note that we do not perform a Voronoi-Tesselation here, because our current implementation of the CPM requires each tile to strictly have six neighbors and thus have a hexagonal shape.
The shape of a hexagon $i$ is defined by its six vertices $\mathbf{v}^{i}_{k}$, which are obtained by interpolating between the positions of hexagon $i$ and two mutually connected nearest neighbors:
\begin{equation}
	\mathbf{v}^{i}_{k} =
	\frac{1}{3}
	\left[
	\mathbf{x}_i + \mathbf{x}_{\mathcal{N}^{i}_{k}} + \mathbf{x}_{\mathcal{N}^{i}_{k-1}}
	\right]
	\, .
\end{equation}
This ensures a circular order in the set of vertices of a hexagon $\mathcal{V}=\{ \mathbf{v}^{i}_{k}\}$ and can be graphically represented as follows:\\
\begin{figure}[H]
	\centering
	\includegraphics{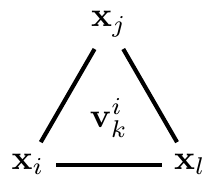}
\end{figure}
Here the hexagons $i$, $j$ and $l$ are pairwise nearest neighbors such that $l \,{=}\, \mathcal{N}^{i}_{k} \,{=}\, \mathcal{N}^{j}_{k+1}$, $j \,{=}\, \mathcal{N}^{i}_{k-1} \,{=}\, \mathcal{N}^{l}_{k+4}$ and $i \,{=}\, \mathcal{N}^{l}_{k+3} \,{=}\, \mathcal{N}^{j}_{k+2}$.

We assume that the hexagons are at all times (including strained states of the substrate) simple polygons.
The area $a$ of a hexagon spanned by the vertices $\mathbf{v}^{i}_{k} \,{=}\, (X^i_k,\, Y^i_k)$ is then given by Gauss area formula 
\begin{equation}
	a(\mathbf{x}_i,t) = \frac{1}{2}\left|\sum_{k=1}^6 X^i_k\left(Y^i_{k+1}-Y^i_{k-1}\right)\right| \,.
\end{equation}
We complete the morphological description of the substrate by defining the six edges $e^{i}_{k}$ of a hexagon $i$ as
\begin{equation}
	e^{i}_{k} = (\mathbf{v}^{i}_{k}, \, \mathbf{v}^{i}_{k+1}) \, ,
\end{equation}
with lengths $|e^{i}_{k}| \,{=}\, |\mathbf{v}^{i}_{k} - \mathbf{v}^{i}_{k+1}|$.
Thus, the edge $e^{i}_{k}$ can also be understood as the border between the hexagons $i$ and $\mathcal{N}^{i}_{k}$.

\subsection{Mathematical description of the cell}
The bulk of the cell is described by a set $\mathcal{C}$ of simply connected hexagons [Fig.~\ref{fig::morphology}]:
\begin{equation}
	\mathcal{C} =
	\Bigl\{
	i \, \bigl| \, i \text{ is occupied by cell}
	\Bigr\} \, .
\end{equation}
The cell membrane is the set of hexagon edges $e^{i}_{k}$ lining the border of the cell bulk $\mathcal{C}$ [Fig.~\ref{fig::morphology}]:
\begin{equation}
	\mathcal{B} =
	\left\{
	e^{i}_{k} 
	\left| \,
	\begin{aligned}
		i \in \mathcal{C} \, , \\
		\mathcal{N}^{i}_{k} \notin \mathcal{C}
	\end{aligned}
	\right.
	\right\}
	\, .
\end{equation} 

\begin{widetext}
\subsection{Observable definitions}
\label{sec::definitions}
In this section we summarise the definition of all observables in Tables.
\centering
\begin{longtable*}[h] { |p{.25\textwidth-2\tabcolsep}|p{.375\textwidth-2\tabcolsep}|p{.375\textwidth-2\tabcolsep}| }
\caption[Cell shape descriptors]{\textbf{Cell shape descriptors}}\\
\toprule
\textbf{Observable} & \textbf{Description and remarks} & \textbf{Definition}\\
\toprule
\endhead
\textbf{Cell area}&
The area of the cell is the sum of the areas of all hexagons occupied by the cell.
&
\begin{equation}
	A = \sum_{i\in\mathcal{C}} a(\mathbf{x}_i,t)
\end{equation}\\
\colrule
\textbf{Cell perimeter}&
The perimeter of the cell is the sum of the lengths of all edges lining the boundary of the cell.
&
\begin{equation}
	P = \sum_{e^{i}_{k}\in\mathcal{B}} |e^{i}_{k}|
\end{equation}\\
\colrule
\textbf{Cell extension}&
The cell shape factor ranges from $0$ (circular cells) to $1$ (infinitely elongated cells). 
&
\begin{equation}
	\alpha = 1-\frac{4\pi A}{P^2}
\end{equation}\\
\botrule
\end{longtable*}

\begin{longtable*}[h] { |p{.25\textwidth-2\tabcolsep}|p{.375\textwidth-2\tabcolsep}|p{.375\textwidth-2\tabcolsep}| }
\caption[Cell position and orientation descriptors]{\textbf{Cell position and orientation descriptors}}\\
%
\toprule
\textbf{Observable} & \textbf{Description and remarks} & \textbf{Definition}\\
\toprule
\endhead
\textbf{Cell coordinates (center of mass)} & 
The center of mass of the cell body is determined under the assumption that each hexagon has the same mass density. & 
\begin{equation} 
	\mathbf{x}_C = \frac{\sum_{i\in \mathcal{C}} a(\mathbf{x}_i) \, \mathbf{x}_i}{\sum_{i\in \mathcal{C}} a(\mathbf{x}_i)}
\end{equation} \\
\colrule
\textbf{Cell coordinates (center of protrusion energy)} &
The center of protrusion energy of the cell body is similar to the center of mass. However, here each hexagon is weighted with its respective protrusion energy. &
\begin{equation}
	\mathbf{x}_\epsilon = \frac{\sum_{i\in \mathcal{C}} \epsilon(\mathbf{x}_i) \, \mathbf{x}_i}{\sum_{i\in \mathcal{C}} \epsilon(\mathbf{x}_i)}
\end{equation} \\
\colrule
\textbf{Cell velocity} &
The cell velocity is obtained from the difference in the center of mass coordinates after $\Delta t \,{=}\, 1 \, \mathrm{MCS}$. &
\begin{equation}
	\mathbf{v}(t) = \frac{\mathbf{x}_C(t+\Delta t) - \mathbf{x}_C (t)}{\Delta t} 
\end{equation} \\
\colrule
\textbf{Instantaneous cell polarization vector} & 
The overall direction of the instantaneous cell polarization always points in the direction of the leading edge of the cell. The superscript~$^\circledast$ indicates the usage of the \textit{non-averaged} (instantaneous) polarization vector &
\begin{equation}
	\mathbf{n}^\circledast_\epsilon = (|\mathbf{n}^\circledast_\epsilon|,\,\theta^\circledast_\epsilon) = \mathbf{x}_\epsilon-\mathbf{x}_C
\end{equation} \\
\colrule
\textbf{Average cell polarization vector} & 
The overall average direction of the cell polarization always points in the direction of the leading edge of the cell. Compared to the instantaneous cell polarization vector $\mathbf{n}^\circledast_\epsilon$, it exhibits less fluctuations. &
\begin{equation}
	\mathbf{n}_\epsilon (t) = (|\mathbf{n}_\epsilon|,\,\theta_\epsilon) = \frac{1}{50} \sum_{t'=0}^{49} \mathbf{n}^\circledast_\epsilon(t+t')
\end{equation} \\
\botrule
%
%
\textbf{Principal axes} & 
The vectors $\mathbf{n}_\pm$ corresponding to the two principal axes of the cell are the eigenvectors of the cell shape covariance matrix $\mathrm{Cov}(\mathcal{C})$; see below for a detailed description. &
\begin{equation}
	\mathrm{Cov}(\mathcal{C}) \, \mathbf{n}_\pm = \lambda_\pm \, \mathbf{n}_\pm
\end{equation} \\
\botrule
\end{longtable*}

\begin{longtable*}[h] { |p{.25\textwidth-2\tabcolsep}|p{.375\textwidth-2\tabcolsep}|p{.375\textwidth-2\tabcolsep}| }
\caption[Cell trajectory descriptors]{\textbf{Cell trajectory descriptors}}\\
\toprule
\textbf{Observable} & \textbf{Description and remarks} & \textbf{Definition}\\
\toprule
\endhead
\textbf{MSD} & 
mean-square Displacement of the cell. &
\begin{equation}
	\big\langle R(t)^2\big\rangle = \big\langle |\mathbf{x}(t_0+t) - \mathbf{x}(t_0)|^2 \big\rangle_{t_0}
\end{equation} \\
\colrule
\textbf{VACF} & 
Normalized Velocity Auto-Correlation Function of the cell. &
\begin{equation}
	C_\mathrm{V}(t) = \left\langle \frac{\mathbf{v}(t_0+t) \, \mathbf{v}(t_0)}{|\mathbf{v}(t_0+t)| \, |\mathbf{v}(t_0)|} \right\rangle_{t_0}
\end{equation} \\
\colrule
\textbf{PACF} & 
Normalized Polarization Vector Auto-Correlation Function of the cell. &
\begin{equation}
	C_\mathrm{P}(t) = \left\langle \frac{\mathbf{n}^\circledast_\epsilon(t_0+t) \, \mathbf{n}^\circledast_\epsilon(t_0)}{|\mathbf{n}^\circledast_\epsilon(t_0+t)| \, |\mathbf{n}^\circledast_\epsilon(t_0)|} \right\rangle_{t_0}
\end{equation} \\
\colrule
\textbf{SAACF} & 
Normalized Short Axis Auto-Correlation Function of the cell. &
\begin{equation}
	C_\mathrm{SA}(t) = \left\langle \frac{\mathbf{n}_{-}(t_0+t) \, \mathbf{n}_{-}(t_0)}{|\mathbf{n}_{-}(t_0+t)| \, |\mathbf{n}_{-}(t_0)|} \right\rangle_{t_0}
\end{equation} \\
\botrule
\end{longtable*}

\begin{longtable*}[h] { |p{.25\textwidth-2\tabcolsep}|p{.375\textwidth-2\tabcolsep}|p{.375\textwidth-2\tabcolsep}| }
\caption[Angular profiles in relative coordinates]{\textbf{Angular profiles in relative coordinates}}\\
\toprule
\textbf{Observable} & \textbf{Description and remarks} & \textbf{Definition}\\
\toprule
\endhead
\textbf{Substrate density (inner cell boundary)} & 
Substrate density at the cell boundary and inside of the cell, at the angle $\theta$ relative to the average direction of cell polarization $\theta_\epsilon$. The relative coordinates are defined as $\tilde{\mathbf{x}} \,{=}\, (\tilde{r},\, \tilde{\theta}) \,{=}\, \mathbf{x} - \mathbf{x}_C$. &
\begin{equation}
	\sigma_\text{b,I}(\theta) = \left\langle \frac{a_0}{a(\mathbf{x}_i)} \right\rangle_{\begin{subarray}{l} e^{i}_{k}\in\mathcal{B},\\ \tilde{\phi}_{i}\approx \theta_\epsilon \pm \theta \end{subarray}}
\end{equation}\\
\colrule
\textbf{Substrate density (outer cell boundary)} & 
Substrate density at the cell boundary and outside of the cell, at the angle $\theta$ relative to the average direction of cell polarization $\theta_\epsilon$. The relative coordinates are defined as $\tilde{\mathbf{x}} \,{=}\, (\tilde{r},\, \tilde{\theta}) \,{=}\, \mathbf{x} - \mathbf{x}_C$. &
\begin{equation}
	\sigma_\text{b,O}(\theta) = \left\langle \frac{a_0}{a(\mathbf{x}_{\mathcal{N}^{i}_{k}})} \right\rangle_{\begin{subarray}{l} e^{i}_{k}\in\mathcal{B},\\ \tilde{\phi}_{\mathcal{N}^{i}_{k}}\approx \theta_\epsilon \pm \theta \end{subarray}}
\end{equation}\\
\colrule
\textbf{Cell protrusion energy (cell boundary)} & 
Cell protrusion energy at the cell boundary, at the angle $\theta$ relative to the average direction of cell polarization $\theta_\epsilon$. The relative coordinates are defined as $\tilde{\mathbf{x}} \,{=}\, (\tilde{r},\, \tilde{\theta}) \,{=}\, \mathbf{x} - \mathbf{x}_C$. &
\begin{equation}
	\epsilon(\theta) = \big\langle \epsilon(\mathbf{x}_i) \big\rangle_{\begin{subarray}{l} e^{i}_{k}\in\mathcal{B}, \\ \tilde{\phi}_{i}\approx \theta_\epsilon \pm \theta \end{subarray}}
\end{equation}\\
\botrule
\textbf{Cell polarization strength} & 
Measure for the strength of the cell polarization, i.e. the distinctness of the cell's edhesion energy profile. &
\begin{equation}
	p = \frac{1}{\pi} \int_0^\pi\mathrm{d}\theta\cos(\theta) \, \epsilon_\text{b}(\theta, \theta_\epsilon) \,\sigma_\text{b,I}(\theta, \theta_\epsilon)
\end{equation}\\
\botrule
\end{longtable*}

\begin{longtable*}[h] { |p{.25\textwidth-2\tabcolsep}|p{.375\textwidth-2\tabcolsep}|p{.375\textwidth-2\tabcolsep}| }
\caption[Two-dimensional profiles in relative coordinates]{\textbf{Two-dimensional profiles in relative coordinates }}\\
\toprule
\textbf{Observable} & \textbf{Description and remarks} & \textbf{Definition}\\
\toprule
\endhead
\textbf{Substrate density profile} & 
The spatial profile of the average substrate density around the average cell polarization axis is obtained by radial and angular binning. The relative coordinates are defined as $\tilde{\mathbf{x}} \,{=}\, (\tilde{r},\, \tilde{\theta}) \,{=}\, \mathbf{x} - \mathbf{x}_C$. &
\begin{equation}
	\sigma(r,\theta) = \left\langle \frac{a_0}{a(\mathbf{x}_i)}\right\rangle_{\tilde{\mathbf{x}}_i\approx(r,\, \theta_\epsilon \pm \theta)}
\end{equation}\\
\colrule
\textbf{Cell occupation probability} & 
The probability of substrate occupation around the average cell polarization axis is obtained by radial and angular binning. Here, $\Theta$ is the Heaviside step function. The relative coordinates are defined as $\tilde{\mathbf{x}} \,{=}\, (\tilde{r},\, \tilde{\theta}) \,{=}\, \mathbf{x} - \mathbf{x}_C$. &
\begin{equation}
	\text{Prob}(r,\theta) = \frac{ \sum_{\tilde{\mathbf{x}}_i\approx(r,\, \theta_\epsilon \pm \theta)} \Theta(\epsilon(\mathbf{x}_i) -q )} { \sum_{\tilde{\mathbf{x}}_i\approx(r,\, \theta_\epsilon \pm \theta)} 1}
\end{equation}\\
\colrule
\textbf{Protrusion energy profile} & 
The spatial profile of the average local protrusion energy around the average cell polarization axis is obtained by radial and angular binning. The relative coordinates are defined as $\tilde{\mathbf{x}} \,{=}\, (\tilde{r},\, \tilde{\theta}) \,{=}\, \mathbf{x} - \mathbf{x}_C$. &
\begin{equation}
	\tilde{\epsilon}(r,\theta) = \big\langle \epsilon(\mathbf{x}_i) \big\rangle_{\tilde{\mathbf{x}}_i\approx(r,\, \theta_\epsilon \pm \theta)}
\end{equation}\\
\colrule
\textbf{Protrusion energy profile (occupied)} & 
The spatial profile of the average local protrusion energy around the cell center and the average cell polarization axis under the condition that the substrate is occupied. &
\begin{equation}
	\epsilon(r,\theta) = \frac{\tilde{\epsilon}(r,\theta)}  {\text{Prob}(r,\theta)}
\end{equation}\\
\colrule
\botrule
\end{longtable*}

\end{widetext}

\subsection{Principal component analysis of the cell}
\label{sec::pca}

We perform a principal components analysis of the cell shape to obtain data on its orientation in the form of its long and short axes $\mathbf{n}_\pm \,{=}\, (|\mathbf{n}_\pm|\, , \theta_\pm)$.
Consider the covariance matrix of the cell, which is defined as
\begin{equation}
	\mathrm{Cov}(\mathcal{C})=
	\left(
	\begin{matrix}
		A_{XX} & A_{XY} \\
		A_{XY} & A_{YY}
	\end{matrix}
	\right) \, .
\end{equation}
With the coordinates of each substrate hexagon relative to the cell center $\tilde{\mathbf{x}}_i \,{=}\, \mathbf{x}_i - \mathbf{x}_C \,{=}\, (\tilde{x}_i \,, \tilde{y}_i)$, the elements of the covariance matrix are given by
\begin{align}
	A_{XX} = \frac{\sum_{i\in \mathcal{C}} a(\mathbf{x}_i) \, \tilde{x}_i \, \tilde{x}_i}{\sum_{i\in \mathcal{C}} a(\mathbf{x}_i)} \, , \\
	A_{XY} = \frac{\sum_{i\in \mathcal{C}} a(\mathbf{x}_i) \, \tilde{x}_i \, \tilde{y}_i}{\sum_{i\in \mathcal{C}} a(\mathbf{x}_i)} \, , \\
	A_{YY} = \frac{\sum_{i\in \mathcal{C}} a(\mathbf{x}_i) \, \tilde{y}_i \, \tilde{y}_i}{\sum_{i\in \mathcal{C}} a(\mathbf{x}_i)} \, .
\end{align}
Then, the short or long axis of the cell is defined as the eigenvector $\mathbf{n}_\pm$ of $\mathrm{Cov}(\mathcal{C})$ corresponding to the smaller or larger eigenvalue $\lambda_\pm$, respectively:
\begin{equation}
	\mathrm{Cov}(\mathcal{C}) \, \mathbf{n}_\pm = \lambda_\pm \, \mathbf{n}_\pm \, .
\end{equation}
Both eigenvectors are chosen such that they point in the direction of the polarization vector: $\mathbf{n}_\pm \cdot \mathbf{n}_\epsilon  \,{>}\,  0$.
Since we are only interested in the direction of the eigenvectors, no particular normalization is needed.

\subsection{Cell persistence measurement}
\label{sec::persistence}

The persistence time of directed migration $\tau$ referenced in the main text denotes a typical timescale on which the cell reorients its direction of migration. 
It is obtained by fitting a persistent random walk model to the mean-square displacement of the cell in the simulations:
\begin{equation}
	\big\langle R(t)^2\big\rangle = 2 v^2 \tau^2 \left[t/\tau + e^{-t/\tau} -1\right] \, ,
\end{equation}
with two fit parameters: $v$ and $\tau$.
Simulations on a deformable substrate are fitted using the Interior Point method~\cite{mathematica,Forsgren:2002}, while reference simulations on a rigid substrate are fitted using the Levenberg-Marquardt method~\cite{mathematica,Marquardt:1963,Levenberg:1944}.

\subsection{Substrate model}

As discussed in section~\ref{sec::substrate_description}, the substrate is described by a triangular lattice with time-dependent nodes $\left\{\mathbf{x}_i(t)\right\}_{i=1,\dots,N}$.
These nodes are elastically coupled with their nearest neighbors by loaded springs of zero rest length and are furthermore subject to a viscous dampening and a traction force $\mathbf{T}$:
\begin{equation}
	\zeta \, \dot{\mathbf{x}}_i = \mathbf{T}(\mathbf{x}_i,t) + k \,\sum_{j\in\mathcal{N}_i}\, (\mathbf{x}_j - \mathbf{x}_i).
\end{equation}
By assuming the rest length of the springs to be zero, we enforce a strictly linear response of the substrate to stresses.
For a different approach of linearizing the full equation of motion including a non-zero spring rest length, we refer the reader to~\cite{Yucht:2013iw}.
We have checked that both approaches yield the same phenomenology.
A second alternative approach would be to use a continuum elastic theory to compute substrate strains.
Note that in the absence of traction forces $\mathbf{T}$, the lattice returns to its 'rest state' (all neighbors $i,j$ have the same distance from each other) due to periodic boundary conditions.
To compute the time-dependent node positions, we use an Euler forward method.

\subsection{Cell model}
\label{sec::model}

\begin{figure*}[t]
	\includegraphics{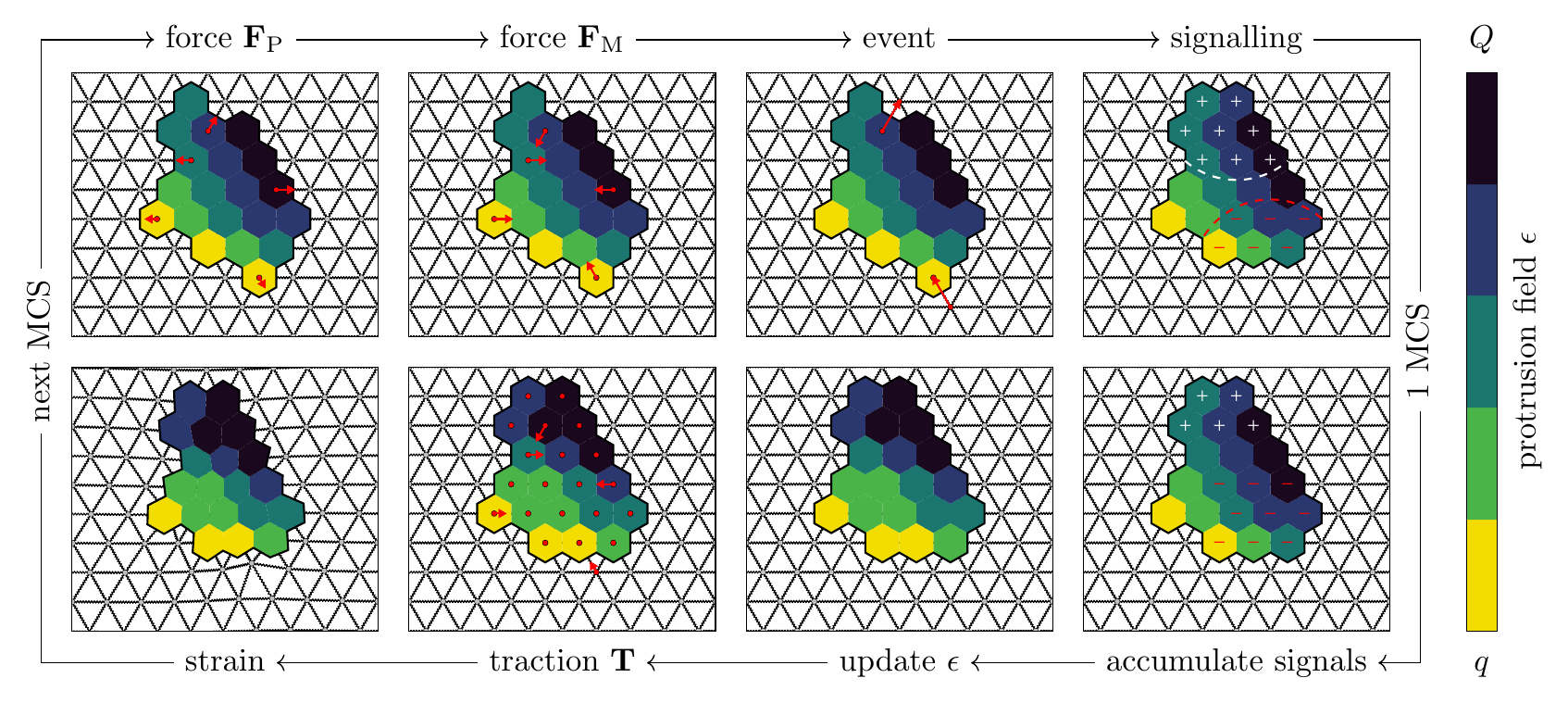}
	\caption{
	\textbf{Overview of a single Monte Carlo Step.}
	An attempted protrusion or retraction event is accompanied by prospective changes in protrusion $\mathcal{H}_P$ and morphological energy $\mathcal{H}_M$.
	These energy changes can be related to effective protrusive $\mathbf{F}_P$ and contractile $\mathbf{F}_M$ forces (illustrated in red for several simultaneously attempted events).
	The acceptance probability of such an event is calculated from the total energy difference $\Delta\mathcal{H}=\Delta\mathcal{H}_P + \Delta\mathcal{H}_M$ [See Eq.~\ref{eq::probability}].
	Successful protrusions are followed by a secretion of internal signals within a radius $R$.
	Similarly, retractions lead to depletion of the mentioned internal signals.
	Over the course of a Monte Carlo Step, many such signals accumulate.
	Then, positive signalling increases the effective local cell protrusion energy $\epsilon$, while negative signalling decreases it.
	Assuming force balance, the protrusive $\mathbf{F}_P$ and contractile forces $\mathbf{F}_M$ can be related to effective traction forces $\mathbf{T}$ on the substrate, leading to deformation. 
	}
	\label{fig::model}
\end{figure*}

For the sake of completeness, we briefly recapitulate the cell model ultilized in this study, which has been previously introduced in \cite{Segerer:2015,Thueroff:2017}.
Please refer to \cite{Segerer:2015,Thueroff:2017} for a detailed discussion and biological motivation of the core model for cell polarity and migration.
For a quick overview over a single Monte Carlo Step, we refer the reader to to Fig.~\ref{fig::model}.

\subsubsection{Metropolis algorithm}

A single Monte Carlo Step in our simulations consists of many individual protrusion or retraction events, where the cell attempts to change its configuration.
By appropriately defining of the total number of protrusion and retraction attempts $|\mathcal{B}|$, we make sure that during a Monte Carlo Step on average each membrane segment will experience contractile [Sec.~\ref{sec::contract}] and protrusive forces [Sec.~\ref{sec::actin}], and as a result attempt to protrude or retract.

During a Monte Carlo Step, a random membrane segment $e^{i}_{k}\in\mathcal{B}$ is selected with a probability proportional to its length:
\begin{equation}
	\text{Prob}(e^{i}_{k}) = \frac{|e^{i}_{k}|}{P} \,.
\end{equation}
With equal probability, the cell attempts to either protrude or retract along the normal vector of the chosen membrane segment $e^{i}_{k}$.
The effective distance vector of such an attempted protrusion is given by $+\mathbf{d}$, while for an attempted retraction it is given by $-\mathbf{d}$ with
\begin{equation}
	\mathbf{d} = \mathbf{x}_{\mathcal{N}_k^i} - \mathbf{x}_i \, ,
\label{eq::distance}
\end{equation}
where $i$ denotes the hexagon inside of the cell that shares the edge $e^{i}_k$ with its $k$-th neighbor $\mathcal{N}^{i}_{k}$ [See Sec.~\ref{sec::substrate_description}].
If a protrusion was successful, the conquered hexagon $\mathcal{N}^{i}_{k}$ is incorporated into the cell bulk
\begin{equation}
	E_+ (e^{i}_{k}): \mathcal{C} \mapsto \mathcal{C} \cup \mathcal{N}^{i}_{k} \, .
\end{equation}
Similarly, in the case of a retraction the hexagon $i$ is removed from the cell bulk
\begin{equation}
	E_- (e^{i}_{k}): \mathcal{C} \mapsto \mathcal{C} \setminus i \, .
\end{equation}
Each cell configuration is associated with a Hamiltonian $\mathcal{H}$.
Thus, changes in the configuration are reflected by the energy state of the cell.
The probability for an event $E_\pm (e^{i}_{k})$ to be successful is then determined by the energy difference $\Delta\mathcal{H}$ between the initial and the attempted cell state
\begin{equation}
	p(\Delta\mathcal{H})=\min\left(e^{-\beta\Delta\mathcal{H}},\,1\right) \,.
\label{eq::probability}
\end{equation}
The inverse effective temperature $\beta$ is a measure for the fluctuations and activity of the cytoskeletal dynamics on a cellular scale.
Thus, in general $\beta$ does not correspond to the room temperature.
The energy difference $\Delta\mathcal{H} \,{=}\, \Delta\mathcal{H}_M + \Delta\mathcal{H}_P$ is determined from the Hamiltonian modelling the contractility of the cell membrane and its cortex ($\mathcal{H}_M$, see Sec.~\ref{sec::contract}), and the Hamiltonian modelling the protrusive actin network ($\mathcal{H}_P$, see Sec.~\ref{sec::actin}).

\subsubsection{Contractility of the cell membrane and cortex}
\label{sec::contract}

The geometry of the cell is constrained by its elastic membrane and the contractile cytoskeleton, which is adhered to the substrate~\cite{Tojkander:2012, Herrmann:2007, Sultan:2004, Tsuruta:2003}.
Thus, it is reasonable to assume in a first approximation that -- similar as in the original CPM~\cite{Graner:1992} -- deformations of a cell's membrane and cortex are constrained by the elastic energy $\mathcal{H}_{M} \,{=}\, \kappa_{A} A(t)^{2} {+} \kappa_{P} P(t)^{2}$ with $\kappa_A$ and $\kappa_P$ denoting the stiffnesses corresponding to the area $A(t)$ and perimeter $P(t)$ of the cell, respectively.
A change in cell morphology is accompanied by a change in the morphological energy $\Delta\mathcal{H}_M$.
This can be related to an effective contractile force always pointing inwards of the cell and acting on the membrane at each attempt to protrude over an effective distance $+\mathbf{d}$ or retract over an effective distance $-\mathbf{d}$ [Eq.~\ref{eq::distance}]:
\begin{equation}
	\mathbf{F}_M(\mathbf{x}_i,t) = -\frac{|\Delta \mathcal{H}_M (\mathbf{x}_i,t)|}{\vert\mathbf{d}(\mathbf{x}_i,t)\vert^2} \, \mathbf{d}(\mathbf{x}_i,t) \, .
\end{equation}
We assume that the cytoskeleton facilitating this contractility transmits forces instantaneously throughout the cell.
Then, the contractile force $\mathbf{F}_M$ is distributed homogeneously over all hexagons $j\in\mathcal{C}$ occupied by the cell and balanced by traction forces $\mathbf{T}_M(\mathbf{x}_j) \,{=}\, {-}\mathbf{F}_M(\mathbf{x}_i)/\vert\mathcal{C}\vert$.
This denotes the traction force contribution stemming from the contractile force on a membrane segment during a single protrusion or retraction event.

Note that the contraction is isotropic throughout the cell and as a result the average contractile forces along the cell membrane $\mathcal{B} \,{=}\, \{e_k^i\}$ vanish 
\begin{equation}
	\left\langle \mathbf{F}_M(\mathbf{x}_i,t) \right\rangle_{e_k^i\in\mathcal{B},\,t} \,{=}\, 0 \, .
\end{equation}
Contractile forces and thus also traction forces resulting from cell contractility are distributed homogeneously over all hexagons occupied by the cell $j\in\mathcal{C}$:
\begin{equation}
	 \left\langle\mathbf{T}_{M}(\mathbf{x}_j)\right\rangle_{t} = \left\langle\mathbf{T}_{M}(\mathbf{x}_j)\right\rangle_{j\in\mathcal{C},\, t} \, .
\end{equation}
Hence, the average traction force contribution on occupied hexagons $j\in\mathcal{C}$ resulting from contractile forces is negligible
\begin{equation}
	\left\langle\mathbf{T}_{M}(\mathbf{x}_j)\right\rangle_{t} = -\left\langle \mathbf{F}_M(\mathbf{x}_i,t)/|\mathcal{C}| \right\rangle_{e_k^i\in\mathcal{B}, \,t} \,{=}\, 0\, .
\end{equation}

\subsubsection{Actin network of the cell}
\label{sec::actin}

The homogeneous contractile forces facilitated by the contractile cytoskeleton are counteracted by local and inhomogeneously distributed outwardly directed pushing forces generated by cytoskeletal structures.
These pushing cytoskeletal structures are locally anchored to the substrate at focal adhesion sites~\cite{Pollard:2003, Mogilner:2009}.
Because of this anchoring, they will behave in an affine way to the substrate, and the local amount of cytoskeleton per hexagon will remain constant under substrate deformations.
Thus, we describe the local energetic contribution from this cellular activity with 
\begin{equation}
	\mathcal{H}_P \,{=}\, {-} \sum_{i\in \mathcal{C}} \epsilon(\mathbf{x}_i,t) \, ,
\end{equation}
with the scalar \textit{protrusion} field \textit{per hexagon} $\epsilon(\mathbf{x}_i,t) \,{\in}\, [q, Q]$~\cite{Thueroff:2017, Segerer:2015}.
The protrusion field is dynamic, reflecting the response of cytoskeletal structures to external mechanical stimuli through feedback mechanisms involving regulatory cytoskeletal proteins~\cite{Maree:2006, Maree:2012}, as will be described in the next section.
Note that this protrusion field could as well be interpreted as a local, inhomogeneously distributed adhesion energy to the substrate.
In this picture, the adhesion energy per hexagon would also remain constant under substrate deformations, as adhesions sites per definition deform affinely with the substrate.

If the cell acquires a new hexagon, the conquered hexagon (target) will have the same protrusion field as the hexagon pushing the membrane (conqueror), and the overall polarization energy increases by the local protrusion field of the conquering hexagon.
This can be interpreted as the pushing cytoskeletal structures moving into the acquired hexagon.
Similarly, in the case of a retraction the overall polarization energy decreases by the local protrusion energy of the lost hexagon.
These energy changes can be related to an effective outward pushing force locally exerted by the cytoskeleton on the cell membrane at each attempt to protrude over an effective distance $+\mathbf{d}$ or retract over an effective distance $-\mathbf{d}$:
\begin{equation}
	\mathbf{F}_P(\mathbf{x}_i,t) = \frac{|\Delta \mathcal{H}_P (\mathbf{x}_i,t)|}{\vert\mathbf{d}(\mathbf{x}_i,t)\vert^2} \, \mathbf{d}(\mathbf{x}_i,t) \, .
\end{equation}
This pushing force is transmitted to the substrate by focal adhesions and balanced locally by a traction force $\mathbf{T}_P \,{=}\, -\mathbf{F}_P$, which points towards the cell interior.

The total traction force stemming from a single protrusion or retraction event  that is locally exerted on the substrate $\mathbf{T} \,{=}\, \mathbf{T}_M + \mathbf{T}_P$ consists of a contribution from the contractility of the cell and and a contribution from the protrusive cytoskeleton.


\subsubsection{Mechanochemical positive feedback}
\label{sec::feedback}

Cell migration is assumed to be driven mainly by a positive feedback loop involving the actin cytoskeleton and some -- a priori unknown -- signalling molecule.
The relative amount of signalling molecules is coarse grained into an integer field $m(\mathbf{x}_i)$, which can also take negative values.
Because the internal dynamics of the cell is assumed to be fast, the amount of signalling molecules is reset after each Monte-Carlo Step. 

Consider a successful protrusion event $E_+(e^{i}_{k})$, with the position of the acquired hexagon given by $\mathbf{y}=\mathbf{x}_{\mathcal{N}^{i}_{k}}$.
Analogously to the protrusion energy, in the case of a successful protrusion the signalling field $m(\mathbf{x}_i)$ of the hexagon facilitating the protrusion is copied unto the acquired hexagon $m(\mathbf{y}) \mapsto m(\mathbf{x}_i)$.
Then, signalling molecules are secreted and diffuse within a signalling radius $R$ of the conquered hexagon:
\begin{equation}
	m(\mathbf{x}_j) \mapsto
	\begin{cases}
		m(\mathbf{x}_j)+1, & \forall j \in \mathcal{C}: |\mathbf{x}_j - \mathbf{y}|<R \\
		m(\mathbf{x}_j), & \text{else.}
	\end{cases}
\end{equation} 
Similarly, in the case of a retraction event $E_-(e^{i}_{k})$ signalling molecules are depleted within the signalling radius $R$ of the lost hexagon $\mathbf{y}=\mathbf{x}_{i}$:
\begin{equation}
	m(\mathbf{x}_i) \mapsto
	\begin{cases}
		m(\mathbf{x}_i)-1, & \forall i \in \mathcal{C}: |\mathbf{x}_i - \mathbf{y}|<R \\
		m(\mathbf{x}_i), & \text{else.}
	\end{cases}
\end{equation}
Because the lost hexagon removed from the cell, its corresponding signalling molecules are reset to zero $m(\mathbf{y}) \mapsto 0$.

These protrusion or retraction events are driven by the actin cytoskeleton, which is modelled by the scalar protrusion energy $\epsilon$.
Protrusion events are more likely to occur in regions of high local protrusion energy $\epsilon$, while retractions are more numerous in regions of low $\epsilon$. 
Throughout a single Monte Carlo Step, many such protrusion and retraction events occur, and the corresponding signals overlap.
Finally, at the end of a Monte Carlo Step with duration $\Delta t$, the actin cytoskeleton is assumed to be reinforced in regions of high protrusive activity, and disassembled in regions of low protrusive activity with a rate $\tilde{g}_\epsilon \,{=}\, g_\epsilon \, \Delta t$
\begin{equation}
	\epsilon(\mathbf{x}_i, t + \Delta t) =
	\begin{cases}
		\epsilon(\mathbf{x}_i, t) + \tilde{g}_\epsilon (Q - \epsilon(\mathbf{x}_i, t)), & m(\mathbf{x}_i) > 0, \\
		\epsilon(\mathbf{x}_i, t) + \tilde{g}_\epsilon (q - \epsilon(\mathbf{x}_i, t)), & m(\mathbf{x}_i) < 0, \\
		\epsilon(\mathbf{x}_i, t) + \tilde{g}_\epsilon (\bar{\epsilon} - \epsilon(\mathbf{x}_i, t)) , & \text{else.}
	\end{cases}
\end{equation}
Thus, a positive feedback loop is incorporated into the Cellular Potts model.
Here, $g_\epsilon$ is a measure for the speed of the cytoskeletal remodelling, and $\bar{\epsilon} \,{=}\, (Q+q)/2$.
Before performing the actual simulations, we pre-equilibrate the cell by letting it grow on a non-deformable substrate for $1000\text{ MCS}$ with the positive feedback switched off and the protrusion field fixed at $\epsilon \,{=}\, \bar{\epsilon}$.
There, the cell starts off as a single hexagon and grows until it reaches equilibrium.


\subsection{Simulation parameters}
\label{sec::parameters}

\begin{table} [t]
	\centering
	\caption[Simulation parameters]{\textbf{Simulation parameters}}
	\begin{tabular} { |p{.25\linewidth-2\tabcolsep}|p{.375\linewidth-2\tabcolsep}|p{.325\linewidth-2\tabcolsep}| }
	\toprule
	\textbf{Parameter} & \textbf{Description} & \textbf{Value(s)}\\
	\botrule
	$\beta^{-1}$ & effective temperature & \SI[per-mode=symbol]{100}{\pico\newton\micro\meter} \\
	\botrule
	& \textbf{Cell} & \\
	$q$ & protrusion energy (lower bound) & \SI[per-mode=symbol]{500}{\pico\newton\micro\meter} \\
	$Q$ & protrusion energy (upper bound) & \SI[per-mode=symbol]{1000}{\pico\newton\micro\meter} \\
	$\kappa_A$ & area stiffness & \SI[per-mode=symbol]{0.5}{\pico\newton\per\micro\meter\cubed} \\
	$\kappa_P$ & perimeter stiffness & \SI[per-mode=symbol]{0.75}{\pico\newton\per\micro\meter} \\
	$R$ & signalling radius of internal cell dynamics & \SI[per-mode=symbol]{7.07}{\micro\meter} \\
	$g_\epsilon$ & update rate of internal cell dynamics & \SI{0.014}{\per\second} \\
	\botrule
	& \textbf{Substrate} & \\
	$k$ & stiffness & \SIrange[range-units = single, per-mode=symbol]{0.5}{8.75}{\nano\newton\per\micro\meter} \\
	$\zeta$ & viscous friction & \SIrange[range-units = single, per-mode=symbol]{17.5}{121}{\second\nano\newton\per\micro\meter} \\
	\botrule
	\end{tabular}
\end{table}

To allow for sufficient ruffling of the cell membrane, we choose the effective temperature $\beta^{-1} \,{=}\, \SI[per-mode=symbol]{100}{\pico\newton\micro\meter}$, which corresponds to an effective temperature much larger than room temperature.
A single cell is simulated over the course of \num{e4} Monte Carlo Steps, each divided into \num{e3} substrate update steps.
We choose the initial distance between adjacent hexagons $d_0=\SI[per-mode=symbol]{1.41}{\micro\meter}$ (lattice constant).
Hence, a cell spreading over an area of \SI[per-mode=symbol]{400}{\micro\meter\squared} consists of roughly \num{2.3e2} hexagons.
The substrate is \SI[per-mode=symbol]{283}{\micro\meter} wide and \SI[per-mode=symbol]{245}{\micro\meter} high, with periodic boundary conditions.

The lower and upper protrusion energy bounds represent the ability of the cell to exert protrusive forces on the membrane and traction on the substrate.
Human umbilical vein endothelial cells have been measured to exert physiological traction stresses up to \SI[per-mode=symbol]{600}{\pascal}~\cite{Pompe:2011}.
On average, similar traction stresses have been measured for fibroblasts, though also reaching up to several \si[per-mode=symbol]{\kilo\pascal}~\cite{Lo:2000}.
Recent measurements have found similar values for the stresses exerted by MDA-MB-231 cells on their three-dimensional environment~\cite{Han:2017}.
To obtain traction forces on the correct order of magnitude, we set $q \,{=}\, \SI[per-mode=symbol]{500}{\pico\newton\micro\meter}$ and $Q \,{=}\, \SI[per-mode=symbol]{1000}{\pico\newton\micro\meter}$ for the lower and upper protrusion energy bounds, respectively.
Assuming the substrate depth to be on the order of the lattice constant \SI[per-mode=symbol]{1.41}{\micro\meter}, the protrusion energy bounds correspond to traction stresses ranging from \SIrange[per-mode=symbol]{306.19}{612.37}{\pascal}.
Similarly, the studied substrate stiffness of \SIrange[per-mode=symbol]{0.5}{8.75}{\nano\newton\per\micro\meter} can be related to an effective elastic modulus ranging from \SIrange[per-mode=symbol]{0.61}{10.72}{\kilo\pascal}.

We study the influence of the viscous friction of the substrate on the cell behavior within the range \SIrange[per-mode=symbol]{17.5}{121}{\second\nano\newton\per\micro\meter}.

To obtain a cell size of approximately \SI[per-mode=symbol]{430}{\micro\meter\squared}~\cite{Riaz:2016}, the area stiffness is chosen as $\kappa_A \,{=}\, \SI[per-mode=symbol]{0.5}{\pico\newton\per\micro\meter\cubed}$.
The low perimeter stiffness $m \,{=}\, \SI[per-mode=symbol]{0.75}{\pico\newton\per\micro\meter}$ allows for significant membrane fluctuations.
For the remaining parameters, we set the signalling radius to $R= \SI[per-mode=symbol]{7.07}{\micro\meter}$ and the polarization update rate $\tilde{g}_\epsilon \,{=}\, g_\epsilon \, \Delta t \,{=}\, 0.1$.

To achieve a cell speed of approximately \SI[per-mode=symbol]{0.1}{\micro\meter\per\second}~\cite{Riaz:2016}, the duration of a single Monte Carlo Step is set to $\Delta t \,{=}\, \SI[per-mode=symbol]{7}{\second}$.

\clearpage

\section{Supplemental discussion}

In our model, the cell can exhibit different migratory states, depending on the mechanical properties of the substrate: \textit{running}, \textit{rounding} and \textit{elongation}.
Complementary to the discussion in the main text, a more extensive explanation of the phenomenology is provided in the sections below.
The additionally provided data serves to improve the intuition for the cell behavior across a wide range of parameters.

\subsection{Measuring the persistence time of directed migration of the cell}
\label{sec::persistence_comparison}

\subsubsection{Mean-square displacement}
\label{sec::MSD}

The persistence time of directed migration $\tau$ represents the typical time over which the cell decorrelates (in other words reorients) its direction of motion.
It has been shown previously that the migration of a polarized cell in the Cellular Potts model~\cite{Thueroff:2017} can be approximated by a persistent random walk model with an exponentially decaying velocity auto-correlation function.
Here we obtain the persistence time of directed migration $\tau$ by fitting the mean-square displacement with the corresponding expression for a persistent random walk model [Section~\ref{sec::persistence}]:
\begin{equation}
	\big\langle R(t)^2\big\rangle = 2 v^2 \tau^2 \left( t/\tau + e^{-t/\tau}-1\right) \, .
\end{equation}
In this model, the cell migrates on an almost straight path on short timescales ($t \ll \tau$) and performs a random walk on long timescales ($t \gg \tau$).

\subsubsection{Normalized velocity auto-correlation}
\label{sec::VACF}

\begin{figure}[t]
	\includegraphics{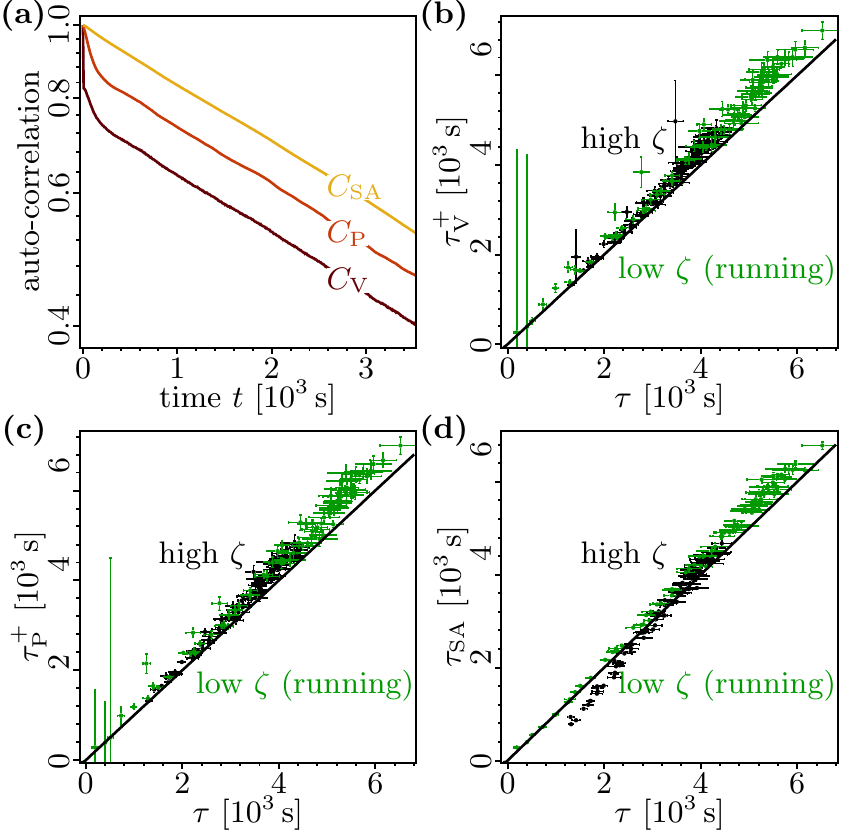}
	\caption{
	\textbf{Correlation functions and persistence times. }
	\textbf{(a)} Exemplary correlation functions for a stiff substrate $k \,{=}\, \SI[per-mode=symbol]{8.75}{\nano\newton\per\micro\meter}$ and low viscous friction $\zeta \,{=}\, \SI[per-mode=symbol]{17.5}{\second\nano\newton\per\micro\meter}$.
	The normalized velocity auto-correlation function (VACF) $C_\mathrm{V}$ exhibits a bi-exponential decay with a long and a short timescale afterwards and a peak at $t \,{=}\, 0$.
	The peak at $t \,{=}\, 0$ can be attributed to the 'randomness' in the protrusion and retraction process (noise).
	In the normalized polarization vector auto-correlation function (PACF) $C_\mathrm{P}$, noise is integrated out, and only the bi-exponential decay remains.
	The normalized short axis auto-correlation function (SAACF) $C_\mathrm{SA}$, which measures the actual reorientation of the cell body, shows a mono-exponential decay (all shorter timescales are integrated out).
	\textbf{(b)-(d)} The timescales obtained by fitting the mean-square displacement (MSD) to a persistent random walk model coincide with the (long) timescales of the VACF, PACF and the SAACF both for low ($\zeta \,{=}\, \SI[per-mode=symbol]{17.5}{\second\nano\newton\per\micro\meter}$) and for high ($\zeta \,{=}\, \SI[per-mode=symbol]{121}{\second\nano\newton\per\micro\meter}$) viscous friction coefficients.
	The error bars correspond to the estimated fitting errors.
	Thus, these are all equivalent measures for the persistence time of directed migration. 
	}
	\label{fig::persistence}
\end{figure}

In, we also investigated the normalized velocity auto-correlation function (VACF) $C_\mathrm{V}$.
Empirically, we find that the behavior is well described by a bi-exponentional decay [Fig.~\ref{fig::persistence}(a)]:
\begin{equation}
	C_\mathrm{V} (t) = a \, e^{-t/\tau_\mathrm{V}^-} + b \, e^{-t/\tau_\mathrm{V}^+} + (1-a-b) \, \delta_{t, 0} \, ,
\label{eq::VACF}
\end{equation}
which is fitted to simulated data using the Interior Point method and four fit parameters $\tau_\mathrm{V}^+  \,{>}\,  \tau_\mathrm{V}^-$ and $b \,{>}\, a$.
We include the spike at $t=0$ to capture the rapid decay of $C_\mathrm{V}$ in the very first time step and to ensure that $C_\mathrm{V}(0) \,{=}\, 1$.
We identify $1-a-b$ as an effective 'noise strength' because the spike at $t=0$ directly originates from the stochasticity of our simulations.
It reflects a 'randomness' in the acceptance of protrusion and retraction events:
Though they are biased by the cell's non-uniform protrusion energy field $\epsilon$, all protrusion and retraction events are essentially stochastic.
In particular, their randomness can be increased either (a) by decreasing the bias resulting from $\epsilon$ or alternatively (b) by increasing the effective temperature $\beta^{-1}$ [Eq.~\ref{eq::probability}].
We observe that the long (dominant) timescale $\tau_\mathrm{V}^+$ coincides with the persistence time of directed migration $\tau$ [Fig.~\ref{fig::persistence}(b)] and thus determines long-term cell behavior.

\subsubsection{Reorientation of the protrusion energy profile}
\label{sec::PACF}

\begin{figure}[b]
	\includegraphics{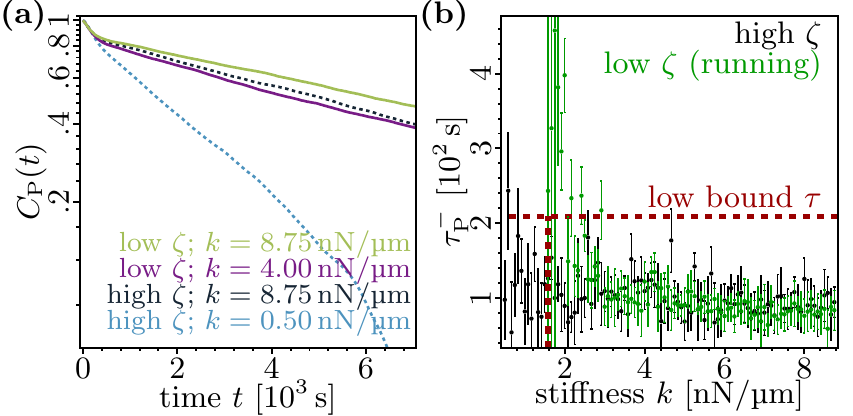}
	\caption{
	\textbf{Polarization vector auto-correlation functions and their short timescale dynamics.}
	\textbf{(a)} Semi-Log plot of exemplary polarization vector auto-correlation functions for different substrate properties as indicated by the graph color.
	At short times $t \,{<}\, \SI[per-mode=symbol]{0.4e3}{\second}$, all correlation functions decay identically.
	\textbf{(b)} The short timescale in the polarization vector auto-correlation function (PACF) $C_\mathrm{P}$ is on the order of \SI[per-mode=symbol]{100}{\second} for different substrate stiffnesses and viscous frictions.
	For low viscous friction ($\zeta \,{=}\, \SI[per-mode=symbol]{17.5}{\second\nano\newton\per\micro\meter}$), the horizontal and vertical dashed lines respectively indicate the lower bound of measured cell persistence times $\tau$ and the corresponding substrate stiffness $k$ [Fig.~\ref{main-fig::stiffness_study}(c)].
	If the long timescale $\tau_\text{P}^+  \,{\approx}\,  \tau$ and the short timescale $\tau_\text{P}^-$ are of the same order of magnitude, the timescale separation in the bi-exponential fit fails.
	The error bars denote the estimated fitting errors.
	}
	\label{fig::shorttau}
\end{figure}

In addition to the mean-square displacement and the normalized velocity auto-correlation function we have also investigated the dynamics of the cell protrusion energy profile and of cell repolarization.
The current orientation of the protrusion energy profile at a given Monte Carlo Step (MCS) is captured by the polarization vector $\mathbf{n}^\circledast_\epsilon \,{=}\, (|\mathbf{n}^\circledast_\epsilon|, \,\theta^\circledast_\epsilon) \,{=}\, \mathbf{x}_\epsilon - \mathbf{x}_C$ pointing from the center of mass $\mathbf{x}_C$ towards the center of protrusion energy $\mathbf{x}_\epsilon$ of the cell [Section~\ref{sec::definitions}].
Here and in later sections, the superscript $^\circledast$ indicates the usage of the \textit{non-averaged} (instantaneous) polarization vector $\mathbf{n}^\circledast_\epsilon$; for the definition of the averaged polarization vector $\mathbf{n}_\epsilon$ we refer the reader to Sec.~\ref{sec::definitions}~and~\ref{sec::averaged_polvector}.
Because protrusions/retractions form preferably in regions of high/low $\epsilon$ respectively, the cell will on average migrate along the gradient of its protrusion energy field $\epsilon$.
Hence the polarization vector pointing towards that side of the cell with a higher local protrusion energy will determine the cell's direction of motion and its leading (protruding) edge.
The change in the instantaneous cell polarization is captured by the normalized polarization vector auto-correlation function (PACF) $C_\mathrm{P}$.
We observe that $C_\mathrm{P}$ exhibits a bi-exponential decay [Fig.~\ref{fig::persistence}(a)]:
\begin{equation}
	C_\mathrm{P} (t) = a \, e^{-t/\tau_\mathrm{P}^-} + (1-a) \, e^{-t / \tau_\mathrm{P}^+} \, ,
\end{equation}
which is fitted using the Interior Point method with three fit parameters $\tau_\mathrm{P}^+  \,{>}\,  \tau_\mathrm{P}^-$ and $a \,{<}\, 0.5$.
The second prefactor $(1-a)$ is determined naturally from the condition $C_\mathrm{P}(0) \,{=}\, 1$.
In contrast to $C_\text{V}$, we do not observe a spike at $t=0$, because the randomness of the protrusion/retraction process is filtered out ('integrated out') by the internal polarization mechanism of the cell.
Because the direction of cell migration is slaved to the direction of instantaneous cell polarization, it is reasonable that both $C_\mathrm{P}$ and $C_\mathrm{V}$ show a similar time evolution and specifically the same decay rates at long timescales [Fig.~\ref{fig::persistence}(b),(c)]. 

We will now illustrate the origin of the short timescale observed in the normalized polarization vector auto-correlation function (PACF).
Let us consider a scenario where the cell is polarized at a given time, that means the cell has a pronounced protrusion energy profile.
Furthermore, we have argued before [Sec.~\ref{sec::VACF}] that there is a certain 'randomness' in the protrusion and retraction processes.
For now let's assume that this 'randomness' dominates and thus the bias of each individual protrusion or retraction event by the local protrusion energy is negligible.
Protrusions and retractions are then in good approximation equally likely everywhere at the cell edge, as would be the case in the limit of high effective temperatures.
In such a scenario, all long-time correlations will be lost to the stochasticity of the cell.
What is then the typical timescale on which a polarized cell will depolarize and change the direction of its polarization vector in response to random protrusion and retraction events?
Such random protrusion and retraction events are filtered by the internal cell dynamics, which is responsible for the formation and maintenance of the cell's protrusion energy profile.
While a stable asymmetric protrusion energy profile can in general not be maintained in the absence of a bias in the protrusion/retraction process (as for e.g. high effective temperatures), it still allows for the formation of a highly volatile transient polarization profile due to the stochasticity of the system.
This transient profile will then decay and its corresponding polarization vector will reorient with a typical timescale set by the internal dynamics of the cell [Section~\ref{sec::parameters}] $\tau_P^- \,{\approx}\, 1/g_\epsilon \,{=}\, \SI[per-mode=symbol]{70}{\second}$.
However, in general the protrusion/retraction process is biased by the protrusion field $\epsilon$, allowing stable polarization profiles.
According to our argumentation, the 'randomness' in the protrusion/retraction process [Sec.~\ref{sec::VACF}] will then typically lead to a small decay and reorientation of the cell's polarization profile on a short timescale set by the internal dynamics of the cell [Section~\ref{sec::parameters}] $\tau_P^- \,{\approx}\, 1/g_\epsilon \,{=}\, \SI[per-mode=symbol]{70}{\second}$.
This is in good agreement with the short timescale $\tau_P^- \,{\approx}\, \SI[per-mode=symbol]{100}{\second}$ that was typically observed in our simulations [Fig.~\ref{fig::shorttau}].
Thus we can conclude that the short timescale $\tau_P^-$ generally observed in our simulations originates from the stochastic behavior of the cell and the resulting decorrelation of the polarization vector at short timescales.

Next let us turn to the origin of the long timescale observed in the normalized polarization vector auto-correlation function (PACF).
For finite effective temperatures cell protrusions and retractions are biased by the cell's protrusion energy field $\epsilon$.
The preference of the cell to protrude at its leading edge and to retract at its trailing edge leads to the reinforcement of the protrusion energy at the leading edge and its weakening at the trailing edge of the cell.
This in turn sustains the protrusion energy profile over long periods of time and leads to the emergence of a long timescale $\tau_\mathrm{P}^+$.

Note that with decreasing substrate stiffness, the persistence time of a cell and thus also the long timescale $\tau_\text{P}^+$ of its PACF decreases.
Once the long timescale becomes small enough to be of the same order as the short timescale, the timescale separation in the bi-exponential fit fails  [Fig.~\ref{fig::shorttau}(b): for low viscous friction $\zeta \,{=}\, \SI[per-mode=symbol]{17.5}{\second\nano\newton\per\micro\meter}$; the corresponding stiffness and persistence time are respectively indicated by the vertical and horizontal dashed lines].
This explains the large spread of $\tau_\mathrm{P}^-$ for low substrate stiffnesses and low viscous friction [Fig.~\ref{fig::shorttau}(b)].

\subsubsection{Reorientation of the cell body}
\label{sec::SAACF}

In the absence of external guiding cues, there are different ways in which a cell can orient itself relative to its direction of migration and vice versa: a symmetrical cell can preferably move along its short axis, along its long axis, or the cell has no particular shape (symmetry) at all and performs a random walk.
In our model the short axis of the cell aligns with the direction of motion and the polarization axis.
Because the internal cell dynamics is much faster than the motion of the cell, the short-timescale decorrelation in the polarization vector auto-correlation function (which arises from random, uncorrelated protrusions or retractions) does not matter for the long-term orientation of the cell and is integrated out.
Thus, if one performs a principal components analysis of the cell shape and considers  the short axis auto-correlation function (SAACF) $C_\mathrm{SA}$, it exhibits a mono-exponential decay [Fig.~\ref{fig::persistence}(a)]:
\begin{equation}
	C_\mathrm{SA} (t) = e^{-t/\tau_\mathrm{SA}} \, ,
\end{equation}
which we fit using the Interior Point method.
In accordance with our arguments, this timescale $\tau_\mathrm{SA}$ coincides with the dominant timescales of the velocity and polarization vector auto-correlation functions, as well as the persistence time $\tau$ [Fig.~\ref{fig::persistence}].
We have thus shown that the long-term behavior of the cell (e.g.~its persistence time of directed migration) can be determined either from the mean-square displacement of the cell, or the velocity, polarization vector or short axis auto-correlation functions [Secs.~\ref{sec::MSD},~\ref{sec::VACF},~\ref{sec::PACF}~and~\ref{sec::SAACF}]. 
Each approach gives quantitatively identical results in a consistent way [Fig.~\ref{fig::persistence}(b)-(d)].

Note that there is a caveat for the measurement of the SAACF: the principal component analysis fails for round cells.
For high viscous friction ($\zeta \,{=}\, \SI[per-mode=symbol]{121}{\second\nano\newton\per\micro\meter}$) cells are rounder than for low viscous friction ($\zeta \,{=}\, \SI[per-mode=symbol]{17.5}{\second\nano\newton\per\micro\meter}$) [Fig.~\ref{main-fig::stiffness_study}] and migrating cells are both rounder and less persistent with decreasing substrate stiffness [Fig.~\ref{main-fig::stiffness_study}].
This explains the deviation of $\tau_\text{SA}$ from $\tau$ for less persistent cells [Fig.~\ref{fig::persistence}(d)].

\subsubsection{Robust measurement of cell orientation}
\label{sec::averaged_polvector}

We have seen that the cell orientation can be captured by using a principal components analysis.
However, this approach fails for round cells because one can then obviously not discern the long from the short axis of the cell.
Hence, the long-term cell orientation (and thus direction of migration) can be robustly measured for all cell shapes only by either using the velocity or the polarization vector, as we know that they both capture long-term cell behavior.
However, the instantaneous polarization vector exhibits short-timescale decorrelations on the order of \SI[per-mode=symbol]{100}{\second} which stem from the intrinsic noise of the Monte-Carlo simulation.
Additionally, the velocity vector is not only slaved to those decorrelations, but furthermore also directly shows the mentioned intrinsic noise.
We therefore choose to utilize the polarization vector for measuring cell orientation, and average it over 50 Monte Carlo Steps (\SI[per-mode=symbol]{350}{\second}): $\mathbf{n}_\epsilon(t) \,{=}\, (|\mathbf{n}_\epsilon|, \, \theta_\epsilon) \,{=}\, \frac{1}{50} \sum_{t'=0}^{49} \mathbf{n}^\circledast_\epsilon(t+t')$.

\subsection{Cell trapping}
\label{sec::oszillations}

\begin{figure}[t]
	\includegraphics{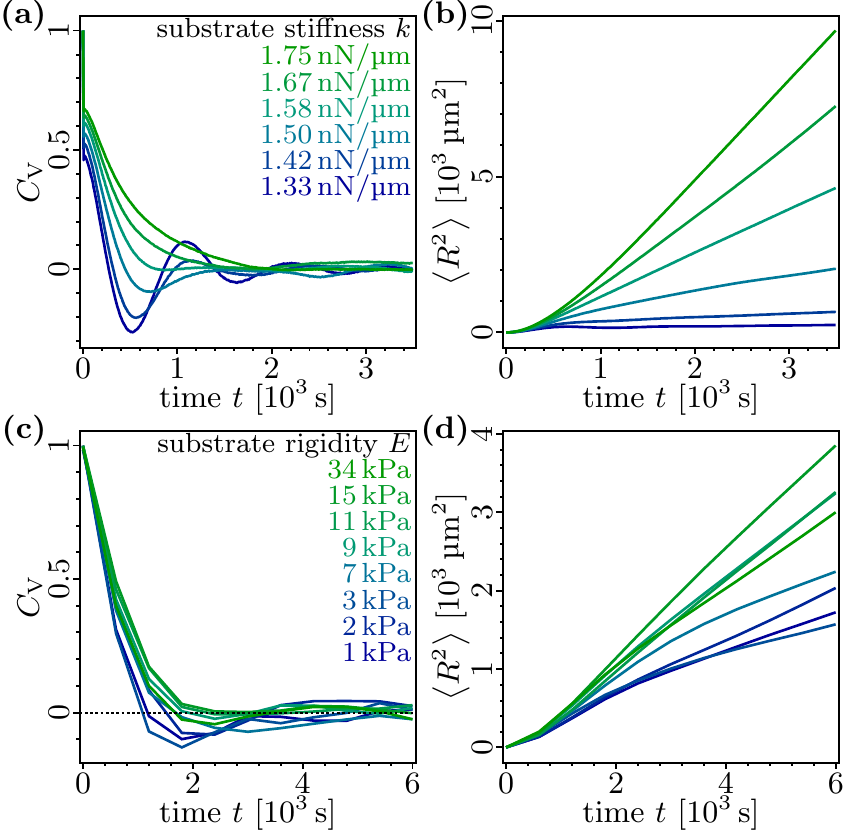}
	\caption{
	\textbf{Persistent motion and self-trapping.}
	\textbf{(a)} Velocity auto-correlation functions $C_\text{V}$ in the simulations for different substrate stiffnesses, as indicated by the graph colors and the legend.
	For $k \,{<}\, k^\star \,{=}\, \SI[per-mode=symbol]{1.58}{\nano\newton\per\micro\meter}$ we observe oszillations in $C_\text{V}$.
	The frequency of these oscillations increases with $k$.
	Note that with increasing $k$ cell migration becomes increasingly uncorrelated.
	\textbf{(b)} Mean-square displacements $\left\langle R^2\right\rangle$ corresponding to the velocity auto-correlation functions shown in (a).
	For $k \,{<}\, k^\star \,{=}\, \SI[per-mode=symbol]{1.58}{\nano\newton\per\micro\meter}$ we observe a saturation of the mean-square displacement.
	In the simulations, the oszillations in $C_\text{V}$ and the saturation of the mean-square displacement strongly indicate cell trapping (cells change from the \textit{running} to the \textit{rounding} state).
	\textbf{(c)} Velocity auto-correlation functions $C_\text{V}$ of HUVECs plated on polyacrylamide (PA) gels for different substrate stiffnesses, as indicated by the graph colors and the legend.
	For $E  \,{<}\,  E^\star  \,{\approx}\,  \SI[per-mode=symbol]{7}{\kilo\pascal}$ we observe anti-correlations in $C_\text{V}$ ($C_\text{V} \,{<}\, 0$).
	\textit{Note:} In the plot we excluded the data point at $\SI[per-mode=symbol]{0.2}{\kilo\pascal}$ because there the cells dramatically changed their mode of migration and ceased the formation of lamellipodia.
	Specifically, for $\SI[per-mode=symbol]{0.2}{\kilo\pascal}$ we observed that cells now migrated preferably along their long axis.
	\textbf{(d)} Mean-square displacements $\left\langle R^2\right\rangle$ of HUVECs plated on PA gels corresponding to the velocity auto-correlation functions shown in (c).
	In our experiments, the anti-correlations in $C_\text{V}$ suggest cell trapping.
	}
	\label{fig::trapping}
\end{figure}

\begin{figure}[t]
	\includegraphics{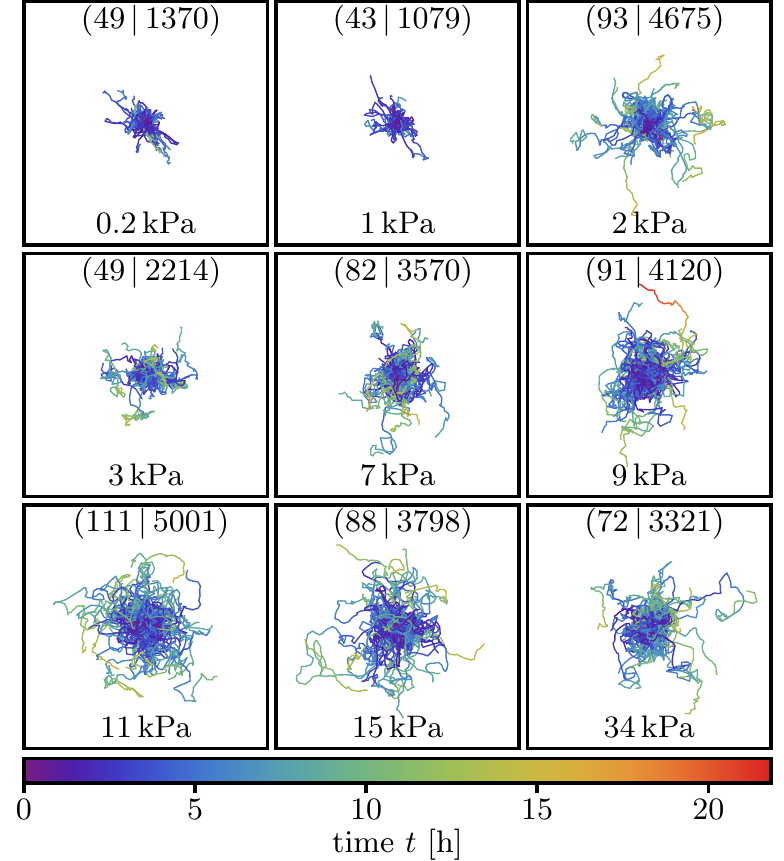}
	\caption{
	\textbf{Cell trajectories in the experiments.} 
	The substrate stiffness is indicated at the bottom of the corresponding frames.
	The amount of measured trajectories $N_T$ and the amount of measured data points $N_P$ is indicated by $(N_T\,|\,N_P)$ at the top of the corresponding frames.
	The color code corresponds to the elapsed time in the respective trajectory (colorbar).
	The duration of the longest measured trajectory is \SI[per-mode=symbol]{21.8}{\hour}.
	Note that the amount of measurements varies for different frames, influencing the visual perception to some extent.
	At low substrate stiffnesses, cells migrate less than at high stiffnesses.
	}
	\label{fig::trajectories}
\end{figure}

In this section, we will briefly discuss the conditions on the mechanical properties of its substrate for the cell to stop performing a persistent random walk and to become self-trapped.
For quickly responding substrates, e.g. low substrate viscous friction, the normalized velocity auto-correlation function (VACF) oscillates if the stiffness falls below a threshold stiffness $k^\star \,{=}\, \SI[per-mode=symbol]{1.58}{\nano\newton\per\micro\meter}$ [Fig.~\ref{fig::trapping}(a)].
Analogously, the mean-square displacement also deviates from that of a persistent random walk model [Fig.~\ref{fig::trapping}(b)], and a typical cell persistence time cannot be determined anymore.
We identify this behavior leading to a decrease in overall cell motility as cell trapping, or, as cell \textit{rounding} because of the corresponding cell shape.

To test our computational results, we have measured the cell trajectories from experiments on HUVECs plated on polyacrylamide gels [Sec.~\ref{sec::experimental_methods}].
We have already seen in the main text that the cell speed decreases with substrate stiffness in qualitative accordance with our model [Fig.~\ref{main-fig::stiffness_study}(a)].
This is complemented by Fig.~\ref{fig::trajectories}, where we can see from the cell trajectories that cell motility increases with substrate stiffness.
Additionally, Fig.~\ref{fig::trapping}(c) shows anti-correlations in the VACF for $E  \,{<}\,  E^\star  \,{\approx}\,  \SI[per-mode=symbol]{7}{\kilo\pascal}$, indicating cell trapping.

\begin{figure}[t]
	\includegraphics{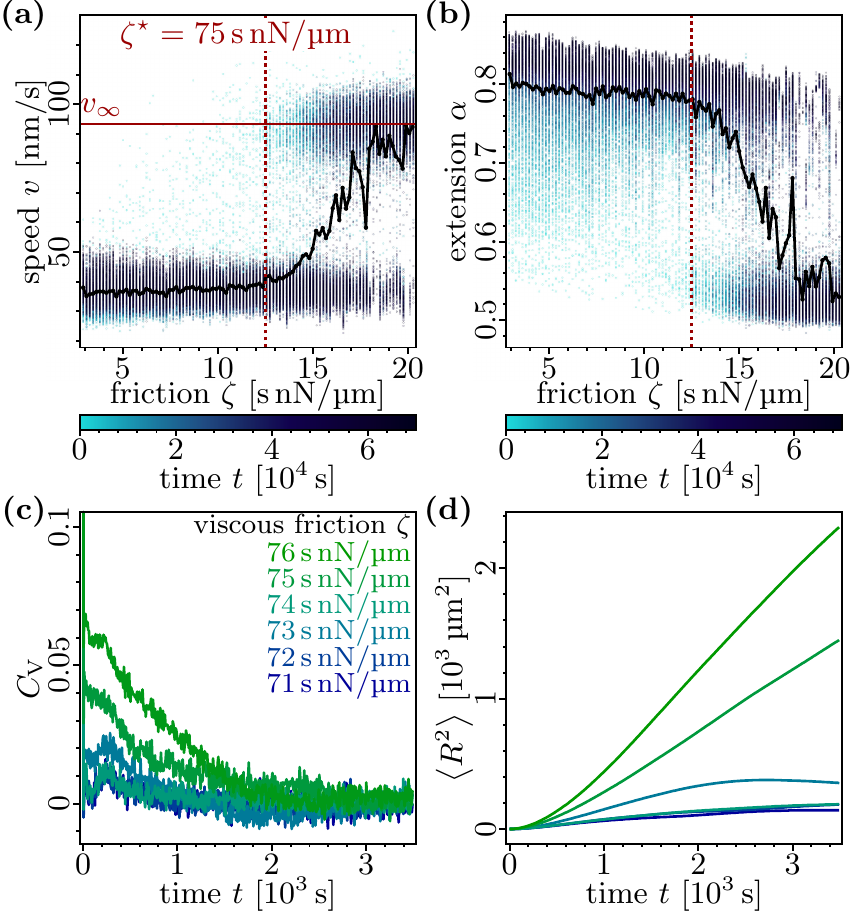}
	\caption{
	\textbf{Dependence of the cell behavior on substrate viscous friction.} 
	\textbf{(a)} Cell speed depends on the viscous friction $\zeta$ of the substrate at low stiffness $k \,{=}\, \SI[per-mode=symbol]{0.5}{\nano\newton\per\micro\meter}  \,{<}\,  k^\star$ [Fig.~\ref{fig::trapping}].
	The color code represents the current elapsed time of a given data point in the simulation (color bar).
	The solid black line corresponds to the averaged behavior of the cells.
	For viscous friction coefficients $\zeta  \,{>}\,  \zeta^\star \,{=}\, \SI[per-mode=symbol]{75}{\second\nano\newton\per\micro\meter}$ indicated by the vertical dashed line, we observe that cells can (at least transiently) migrate and elude trapping.
	Migrating cells then have a typical velocity $v  \,{\approx}\,  v_\infty$, where the horizontal red line denotes the velocity $v_\infty$ of a cell on a non-deformable substrate.
	\textbf{(b)} Cell extension depends on the viscous friction $\zeta$ of the substrate at low stiffness.
	The color code represents the current elapsed time of a given data point in the simulation (color bar).
	The solid black line corresponds to the averaged behavior of the cells.
	For $\zeta  \,{>}\,  \zeta^\star$ indicated by the vertical dashed line we observe that cells can (at least transiently) drastically decrease their elongation.
	Together with (a) this suggests that the cells switch from the \textit{elongation} to the \textit{running} state.
	\textbf{(c)}	Velocity auto-correlation functions $C_\text{V}$ in the simulations for different substrate viscous friction $\zeta$, as indicated by the graph colors and the legend.
	Note the dramatic decay of $C_\mathrm{V}$ from $C_\mathrm{V} \,{=}\, 1$ to $C_\mathrm{V}  \,{\approx}\,  0.05$ in the first time step.
	For $\zeta  \,{>}\,  \zeta^\star \,{=}\, \SI[per-mode=symbol]{75}{\second\nano\newton\per\micro\meter}$, long range correlations in the velocity auto-correlation appear and the functional form of the mean-square displacement of the cell approaches that of a persistent random walk model.
	\textbf{(d)} Mean-square displacements $\left\langle R^2\right\rangle$ corresponding to the velocity auto-correlation functions shown in (c).
	}
	\label{fig::viscosity_detailed}
\end{figure}

We have seen in the main text that for low stiffness and low viscous friction of the substrate, cells have marginal motility due to trapping, and that motility can be restored if the substrate stiffness is high enough [Fig.~\ref{main-fig::stiffness_study}].
Can a similar effect be induced by tuning the viscous friction of the substrate?
To answer this question, we have performed simulations at a fixed low substrate stiffness $k \,{=}\, \SI[per-mode=symbol]{0.5}{\nano\newton\per\micro\meter}  \,{<}\,  k^\star$, where cells are trapped and in the \textit{elongation} state for low viscous friction $\zeta$, and have varied $\zeta$.
Our measurements clearly indicate that with increasing viscous friction of the substrate an increasing amount of cells exhibit persistent cell migration [Fig.~\ref{fig::viscosity_detailed}(a),(b)].
Note that an individual trapped cell is characterized by an oscillating normalized velocity auto-correlation function (VACF) and a saturating mean-square displacement (MSD), while a migrating cell is characterized by a bi-exponentially decaying VACF and no such saturation in the MSD.
In particular, note that the VACF decays faster with decreasing substrate stiffness [Fig.~\ref{fig::trapping}].
Furthermore, we cannot assume that at $\zeta  \,{\approx}\,  \zeta^\star$ \textit{all} cells will be able to outrun the substrate deformations and elude trapping at \textit{all} times; due to the stochastic nature of cell migration a cell might just reorient itself and become trapped.
In general the normalized velocity auto-correlation function [Fig.~\ref{fig::viscosity_detailed}(c)] and the mean-squared displacement [Fig.~\ref{fig::viscosity_detailed}(d)] averaged over all measured cells will then be a weighted average of both trapped and migrating cells.
We expect that with increasing viscous friction of the substrate the contribution of migrating cells to the VACF and to the MSD increases.
Our expectation is confirmed by the increase in correlated cell movement [Fig.~\ref{fig::viscosity_detailed}(c)] and by the MSD \textit{de}saturation [Fig.~\ref{fig::viscosity_detailed}(d)] for high viscous friction coefficients $\zeta  \,{>}\,  \zeta^\star \,{=}\, \SI[per-mode=symbol]{75}{\second\nano\newton\per\micro\meter}$.

Why can, for high substrate viscous friction, cell migration occur even at low substrate stiffnesses?
For high enough viscous friction, the response of the substrate ($\tau_\text{R} \,{=}\, \zeta / k$) is slow compared to the cell dynamics ($\tau_\mathcal{C} \,{\geq}\,  1/g_\epsilon$), and the cell will approximately behave as on a locally non-deformable substrate.
Then, the cell will polarize strongly even on low substrate stiffnesses, before the positive feedback loop can be significantly inhibited by substrate deformations.
By comparing the corresponding timescales of substrate response and cell response ($\tau_\text{R} \,{=}\, \zeta/k \,{\geq}\, \tau_\mathcal{C} \,{\geq}\,  1/g_\epsilon $) we can make a simple estimation that viscous effects become dominant above a lower bound of viscous friction of at least $\zeta^\star \,{\geq}\, \SI[per-mode=symbol] {35}{\second \nano\newton\per\micro\meter}$.
Indeed, it can be observed that $\zeta^\star \,{=}\, \SI[per-mode=symbol] {75}{\second \nano\newton\per\micro\meter}$ [Fig.~\ref{fig::viscosity_detailed}].
This indicates a parameter regime where the cell is fast enough to 'outrun' substrate deformations.
However, due to the stochastic nature of our simulations, it cannot be expected that the cell continues to do so indefinitely.
Specifically, the cell can still become trapped if it runs into a localized region of strongly increased substrate density, e.g. by making a u-turn and running into its trail of increased substrate density [Fig.~\ref{fig::states_simulation}(e)-(h)].
Thus, we expect that even for high viscous friction, we will sporadically observe trapped cells if we wait long enough [Fig.~\ref{fig::viscosity_detailed}(a),(b)].

\subsection{Cell migration on a deformable substrate}

\begin{figure*}[t]
	\includegraphics{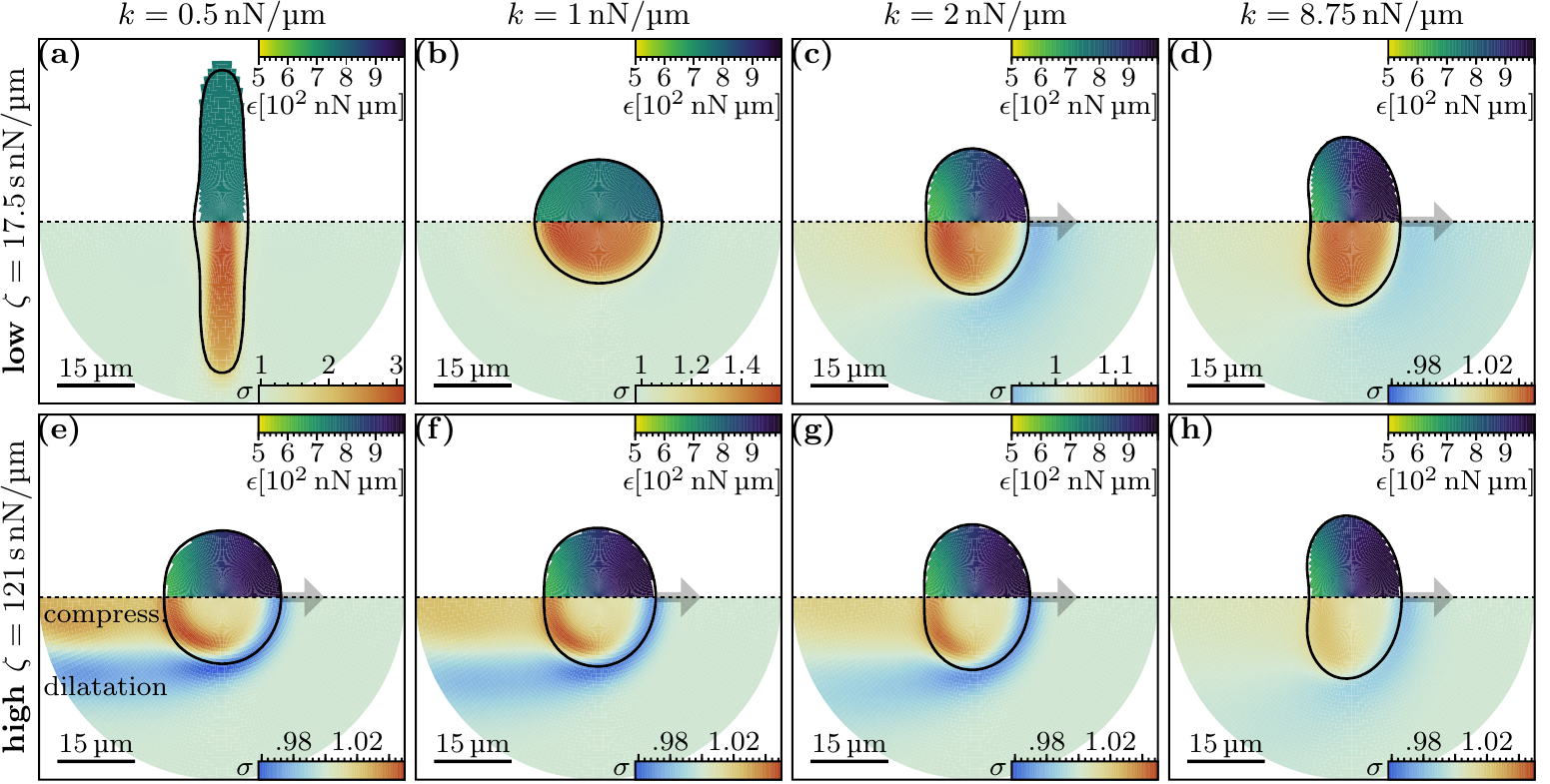}
	\caption{
	\textbf{Cell polarisation and substrate deformation.}
	The top halves and bottom halves of each panel represent the two mirrored halves of a cell.
	The bottom half of each panel depicts the local protrusion energy of the cell \textit{per hexagon} $\epsilon(x,y)$.
	The bottom half of each panel depicts the substrate density below the cell (number of hexagons per unit area) $\sigma(x,-y)$.	
	These quantities are obtained by collapsing the data of many cells in their center of mass frame with the polarization axis oriented along the $x$-axis ($\theta_\epsilon \,{=}\, 0$).
	In general substrate is dilated outside of the cell and compressed inside of the cell.
	Note the differences in the scales of the substrate density $\sigma$.
	The cell will preferably migrate along the gradient $\nabla(\epsilon \, \sigma)$.
	\textbf{(a)},\textbf{(b)} For large substrate deformations, the migration of the cell is in the leading order dominated by the gradient in substrate density $\nabla\sigma$, thus trapping the cell in the region of high substrate density.
	\textbf{(c)-(h)} For small substrate deformations, cell migration is in the leading order dominated by the gradient in protrusion energy $\nabla\epsilon$ and will migrate in the direction indicated by the black arrows.
	}
	\label{fig::states_simulation}
\end{figure*}

\begin{figure*}[t]
	\includegraphics{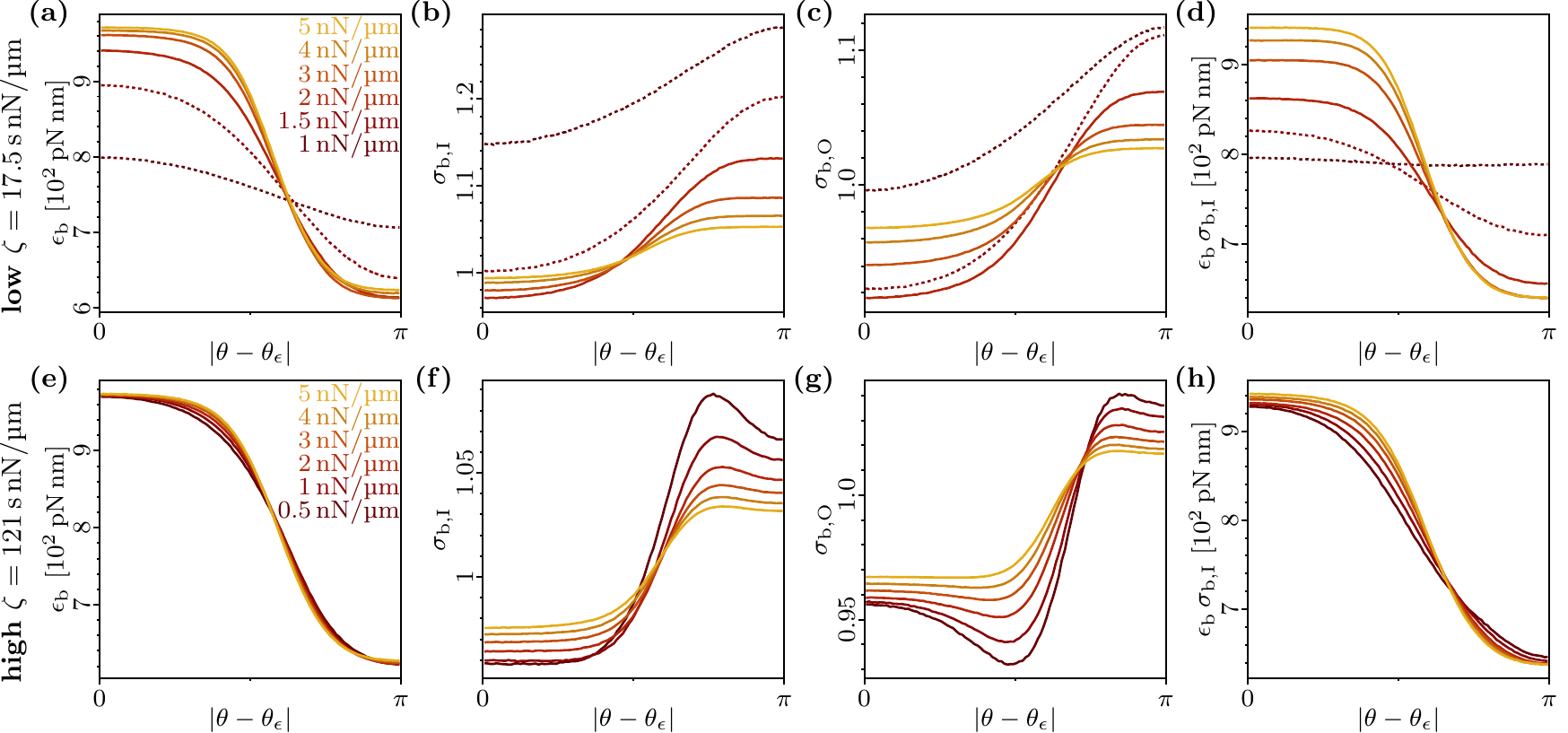}
	\caption{
	\textbf{Angular profiles of cell polarisation and substrate deformation.}
	\textbf{(a),(e)} Profile of protrusion energy \textit{per hexagon} at the boundary of a cell $\epsilon_\text{b}$ for low ($\zeta \,{=}\, \SI[per-mode=symbol]{17.5}{\second\nano\newton\per\micro\meter}$) and high ($\zeta \,{=}\, \SI[per-mode=symbol]{121}{\second\nano\newton\per\micro\meter}$) viscous friction, respectively.
	\textbf{(b),(f)} Profile of substrate density (number of hexagons per unit area) along the interior boundary of a cell $\sigma_\text{b,I}$ for low and high viscous friction, respectively.
	\textbf{(c),(g)} Profile of substrate density (number of hexagons per unit area) along the exterior boundary of a cell $\sigma_\text{b,O}$ for low and high viscous friction, respectively.
	\textbf{(d),(h)} Profile of protrusion energy density at the boundary of the cell $\epsilon_\text{b} \, \sigma_\text{b,I}$ for low and high viscous friction, respectively.
	The dashed lines correspond to cells in the \textit{rounding} state [Fig.~\ref{fig::trapping}: cells are trapped and have an oscillating VACF], while solid lines denote cells in the \textit{running} state.
	}
	\label{fig::distributions}
\end{figure*}

In this section, we present a few exemplary cases for both low viscous friction ($\zeta \,{=}\, \SI[per-mode=symbol]{17.5}{\second\nano\newton\per\micro\meter}$) and high viscous friction ($\zeta \,{=}\, \SI[per-mode=symbol]{121}{\second\nano\newton\per\micro\meter}$) with the intent to give the reader a better intuitive understanding of the dynamics underlying both cell migration and substrate deformation in our Cellular Potts model.
Figure~\ref{fig::states_simulation} shows the two-dimensional spatial profiles of protrusion energy and substrate density.
The corresponding angular profiles of protrusion energy, substrate density and protrusion energy density are shown in Fig.~\ref{fig::distributions}.

\subsubsection{Cell migration on a stiff substrate}
\label{sec::migration_nondeformable}

To obtain a fitting explanation for a new phenomenology, it is often fruitful to begin by considering a simplified and particular example, before advancing to the more general and complex case.
Hence, we will first describe cell migration for vanishingly small substrate deformations in the limit of large substrate stiffness $k  \,{\rightarrow}\,  \infty$ or large viscous friction $\zeta  \,{\rightarrow}\,  \infty$.
Then, all hexagonal substrate tiles have the same size and shape, resulting in a uniform substrate density $\sigma \,{\equiv}\, 1$.
This line of argument can as a first approximation also be applied to simulations showing negligible substrate deformations [Fig.~\ref{fig::states_simulation}(d)-(h)].

The cell makes random protrusions or retractions at its boundary, as described in detail in section~\ref{sec::model}.
A high local protrusion energy $\epsilon$ increases the rate of making protrusions relative to the rate of making retractions, by increasing the energy gain/loss for protrusions/retractions, respectively.
Although the direction of cell migration on average aligns with the cell's protrusion energy field, the core of the model is still stochastic.
Because of this stochastic nature of the computational model, the outcome of an event is a priori unknown, and only the relative probabilities of protrusions or retractions are biased by the local protrusion energy $\epsilon$.
This means that, though less likely, retractions can also occur at the leading edge, and protrusions can also occur at the trailing edge of the cell.

Now, let us consider a scenario with a more pronounced protrusion energy profile, e.g.~the gradient in protrusion energy is steeper than before.
This will lead to an increased bias for protrusions to form at the leading edge and retractions at the trailing edge, and thereby to a better alignment of the cell's individual protrusions and retractions with its protrusion field.
Hence, we can conclude that the alignment of the cell velocity $\mathbf{v} \,{=}\, (v, \, \theta_v)$ with its instantaneous polarization axis $\mathbf{n}^\circledast_\epsilon \,{=}\, (|\mathbf{n}^\circledast_\epsilon|, \, \theta^\circledast_\epsilon)$ will also improve, and the overall effect of stochasticity on cell behavior will decrease.
Furthermore, as it takes more effort to rotate a pronounced protrusion energy profile than to rotate a flat protrusion energy profile, we would expect an increased persistence time of directed migration.
Observations in agreement with our qualitative assessment have been made in the preceding study~\cite{Thueroff:2017} by increasing the maximal polarizability of the cells $\Delta Q \,{=}\, Q-q$.

\subsubsection{Can the cell shape serve as memory?}

We cannot exclude the hypothesis that cell shape contributes to the memory of the cell, in addition to its protrusion energy field.
Because we have assumed that the internal cell dynamics of the cell is much faster than individual protrusions/retractions, reorientations of the cell's protrusion energy field $\epsilon$ should in principal be faster than rotations of the whole cell body.
However, we have observed that cells on substrates with stiffnesses \SI[per-mode=symbol]{8}{\nano\newton\per\micro\meter} and \SI[per-mode=symbol]{4}{\nano\newton\per\micro\meter} have different persistence times of directed migration in spite of having similar shapes [Fig.~\ref{main-fig::stiffness_study}(b),(c)].
Thus, we conclude that cell shape can not be the sole originator of the cell memory.

\subsubsection{How are substrate deformations induced?}

Because the total mass of the substrate is conserved under deformations, a substrate dilatation (decrease in substrate density) at any given location has to be accompanied by a substrate compression (increase in substrate density) elsewhere.
Furthermore, cell traction forces are always pointed inwards of the cell.
Together, this imposes substrate density gradients with substrate dilatation at the cell rim and substrate compression at the cell center, both proportional to the applied traction force.
At the leading edge of the cell, traction forces are larger than at the trailing edge, due to a higher local protrusion energy [Figs.~\ref{fig::states_simulation}~and~\ref{fig::distributions}(a),(e)].
This leads to a stronger substrate dilatation at the leading edge than at the trailing edge of the cell, and a net increase of substrate density at the trailing edge [Figs.~\ref{fig::states_simulation}~and~\ref{fig::distributions}(b),(c),(f),(g)]. 

\subsubsection{What determines cell speed?}

As we have discussed in sections~\ref{sec::VACF}~and~\ref{sec::PACF} the 'noise strength' is a measure for the randomness of cell protrusions and retractions at the cell boundary.
This randomness can be increased by increasing the effective temperature $\beta$ or analogously by decreasing the bias introduced by the local protrusion energy $\epsilon$.
In general, unlike in section~\ref{sec::migration_nondeformable}, the substrate can be deformed.
Thus, the probability of locally gaining or losing an infinitesimal area $dA$ is determined by the local protrusion energy \textit{density} $\epsilon \, \sigma$.
A stronger random contribution in the protrusion/retraction process leads directly to a broader cell velocity distribution around the instantaneous cell polarization vector and consequently a lower cell speed.
This can easily be illustrated as follows:
Consider a cell starting with a pronounced polarization profile (i.e. no initial symmetry breaking required).
Then, in the low temperature limit ($\beta  \,{\rightarrow}\,  \infty$), the cell will \textit{always} protrude at the location of highest polarization energy density and \textit{always} retract at the location of the lowest polarization energy density, leading to a narrow cell velocity distribution.
Because each individual protrusion and retraction event displaces the cell in the same direction and the total amount of such events per Monte Carlo Step is limited, this will lead to a high cell translocation speed.
Conversely, in the high temperature limit ($\beta  \,{\rightarrow}\,  0$), cell protrusions are completely unbiased by the polarization profile, leading to a flat cell velocity distribution.
Because all individual protrusion and retraction events displace the cell in different directions and the total amount of such events per Monte Carlo Step is limited, this will lead to a negligible cell translocation speed.

\begin{figure}[t]
	\includegraphics{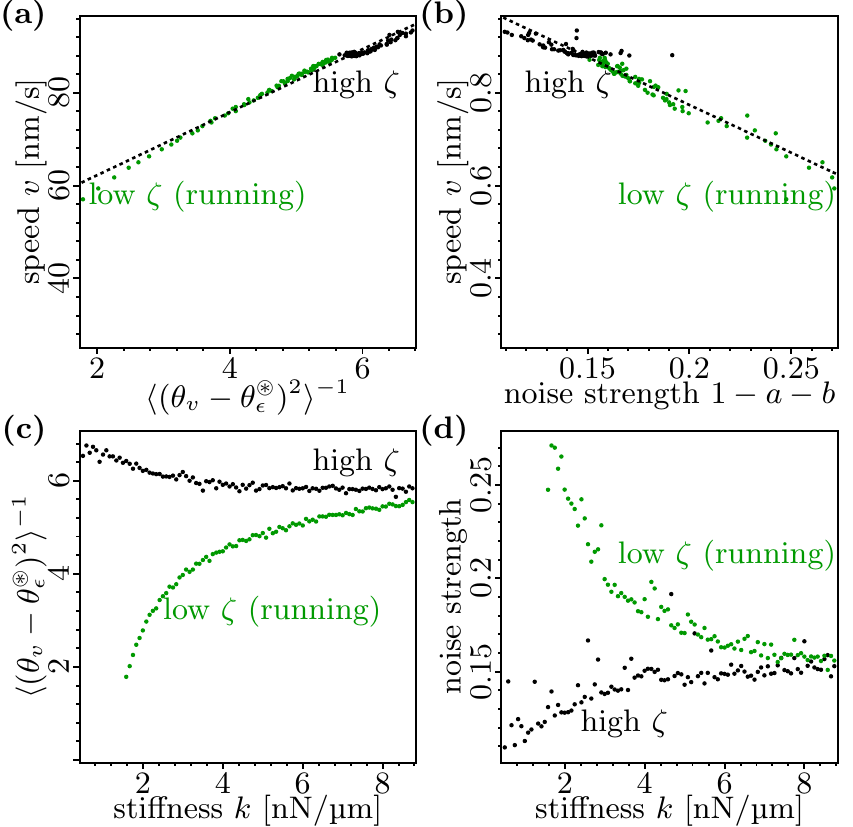}
	\caption{
	\textbf{Stochasticity of the protrusion/retraction process in the simulations.}
	\textbf{(a)} The cell speed decreases with the distribution width of the velocities around the polarization axis of the cell.
	Here, $\theta_v - \theta^\circledast_\epsilon$ is the angle between the cell velocity vector $\mathbf{v}$ and the instantaneous cell polarization vector $\mathbf{n}^\circledast_\epsilon$. 
	\textbf{(b)} The cell speed decreases with increasing 'noise strength' (stochasticity of the protrusion/retraction process) inferred from Eq.~\ref{eq::VACF}. 
	\textbf{(c)} Dependence of the distribution width of the velocities around the polarization axis of the cell on the substrate stiffness $k$.
	\textbf{(d)} Dependence of the 'noise strength' inferred from Eq.~\ref{eq::VACF} on the substrate stiffness $k$.
	}
	\label{fig::noise_speed}
\end{figure}

We conclude that the cell will migrate faster if the bias for individual protrusion/retraction events to align with the cell's protrusion energy field is increased.
Furthermore, this alignment can be measured by the width of the symmetrical velocity distribution around the polarization axis, or the overall randomness of the protrusion/retraction process ('noise strength') inferred from Eq.~\ref{eq::VACF}.
We consistently find that the cell speed increases with both decreasing cell velocity distribution width and with decreasing 'noise strength' [Fig.~\ref{fig::noise_speed}].

\subsubsection{What determines the persistence time of directed migration of the cell?}

To answer this question, we introduce a quantity that characterizes the polarization strength of a cell 
\begin{equation}
	p = \frac{1}{\pi} \int_0^\pi\mathrm{d}\theta\cos(\theta) \, \epsilon_\text{b}(\theta,\theta_\epsilon) \,\sigma_\text{b,I}(\theta,\theta_\epsilon) \, ,
\label{eq::pol_strength}
\end{equation}
with the angle $\theta_\epsilon$ of the average polarization axis relative to the $x$-axis.
The choice of this quantity is based on the idea that the cell is strongly polarized, if there are many tiles (high substrate density) with a high protrusion energy at the leading edge of the cell, and few tiles (low substrate density) with low protrusion energy at the trailing edge of the cell.
The persistence time of directed migration increases exponentially with the cell polarization strength, because it is more costly to rotate the polarization vector of a cell with a pronounced protrusion energy profile than that of an unpolarized cell [Fig.~\ref{main-fig::stiffness_study}(d)].

\subsubsection{Low viscous friction}

For low viscous friction, substrate relaxation and response to the traction forces of the cell occur on short time scales $\tau_\text{R} \,{=}\, \zeta / k$ compared to the typical timescale of the cell dynamics $\tau_\mathcal{C}$.
Thus substrate deformations keep up with the cell and -- depending on the substrate stiffness $k$ -- can become large enough to impair cell motion.

For high values of substrate stiffness [Fig.~\ref{fig::states_simulation}(d)], substrate deformations are small due to high restoring forces.
Thus, the influence of the substrate on the cell behavior becomes negligible and the cell behaves similarly as on a non-deformable substrate ($k \,{\rightarrow}\, \infty$), see section~\ref{sec::migration_nondeformable}.
The cell polarizes strongly and has a long persistence time of directed migration.
Moreover, the cell velocities are narrowly distributed around the cell polarization axis, in correspondence with the high cell speed.

In contrast, low substrate stiffness [Fig.~\ref{fig::states_simulation}(b)] leads to a quick and profound compression of substrate at the position of the cell [Fig.~\ref{fig::distributions}(b)] due to the low restoring forces.
To effectively translocate, the cell needs to protrude at one side and retract at the opposite side.
However, because the substrate density is dramatically increased at the position of the cell, all retractions are energetically penalized, and it becomes energetically disadvantageous for the cell to translocate.
Analogously, one might consider the event that the cell, due to the stochastic nature of the model, manages to move in some random direction.
In that case, it would be energetically advantageous to simply move back to its previous location.
Because the positive feedback mechanism is fueled by protrusions \textit{and} retractions alike [Section~\ref{sec::model}], inhibiting retractions effectively inhibits the formation of a pronounced cell protrusion energy density profile [Fig.~\ref{fig::distributions}(d)].
As a result, there is no notable bias towards protrusions or retractions throughout the cell, leading to a broad velocity distribution and consequently a low cell speed.
Furthermore, the flat energy density profile can be related to a marginal cell polarization strength, thus making cell reorientations cheap and frequent.
Altogether, this leads to an effective cell trapping, where the cell initially attempts to polarize and is then prompted to turn around in its attempt to occupy areas of high substrate density.
This is evidenced by the oscillating VACF [Section~\ref{sec::oszillations}] and leads to a vanishing cell persistence time of directed migration.

The dynamics of cells seeded at intermediate substrate stiffnesses can be understood as an interpolation between the low and high stiffness cases:
in Fig.~\ref{fig::states_simulation}(c), the substrate stiffness is slightly increased compared to Fig.~\ref{fig::states_simulation}(b).
This reduces substrate deformations and leads to less inhibition of the positive feedback mechanism as compared to Fig.~\ref{fig::states_simulation}(b).
Thus, here the cell can establish a more pronounced protrusion energy density profile than in Fig.~\ref{fig::states_simulation}(b) and migrate persistently, i.e.~no oscillations occur in the VACF.
The width of the protrusion energy density profile, and consequently also the cell speed, lies in between Fig.~\ref{fig::states_simulation}(b) and Fig.~\ref{fig::states_simulation}(d).

\subsubsection{High viscous friction}

For high enough viscous friction [Fig.~\ref{fig::states_simulation}(e)-(h)], substrate deformations are too small to completely inhibit the positive feedback mechanism.
Hence, the cell can always polarize and migrate persistently, similar to Fig.~\ref{fig::states_simulation}(d).
The profile of protrusion energy $\epsilon$ \textit{per hexagon} does not particularly change for different stiffnesses [Fig.~\ref{fig::distributions}(e)], and can thus by itself not explain the observed noticeable change in cell persistence time of directed migration.
This can be corrected by taking the substrate density $\sigma$ into account, which represents the \textit{number of hexagons per area}, and thus considering the protrusion energy density $\epsilon \, \sigma$ [Fig.~\ref{fig::distributions}(h)].
The total polarization strength is represented by $p$ [Eq.~\ref{eq::pol_strength}].

Let us first compare the behavior of a cell on high stiffness [Fig.~\ref{fig::states_simulation}(h)], high viscous friction substrate to that on a high stiffness, low viscous friction substrate [Fig.~\ref{fig::states_simulation}(d)].
The cell speeds have similar magnitude due to a similar width of the velocity distribution.
Cell persistence, however, is lower in Fig.~\ref{fig::states_simulation}(h), because the particular substrate density profile effectively reduces the cell polarization strength $p$ in a corresponding way.
In particular, note that the substrate density is reduced at the leading edge, and increased at the trailing edge.

For low substrate stiffness [Figs.~\ref{fig::states_simulation}(e)~and~\ref{fig::distributions}(g)], the lowest substrate density is not encountered at the leading edge, but at the side of the cell.
This particular substrate density profile decreases the probability to protrude at the side of the cell, compared to protruding at the leading edge.
Because the cell has a strong bias to protrude at the leading edge, and the particular substrate density profile slightly discourages motion to the side, the cell velocity distribution around the polarization vector is focussed.
This effectively increases the average cell speed compared to Fig.~\ref{fig::states_simulation}(d), where no such focussing takes place.
Compared to Fig.~\ref{fig::states_simulation}(d), the particular substrate density profile further reduces the cell polarization strength $p$, and thus the persistence time of directed migration, because substrate deformations are larger on a soft substrate than on a stiff substrate.

As before, intermediate substrate stiffnesses can be understood as an interpolation between the low and the high stiffness cases [Fig.~\ref{fig::states_simulation}(f),(g)].

\subsection{Modulation of cell morphology by substrate interactions}

The change in cell shape due to substrate deformations can be explained in a simple way by looking at the substrate density.
In general, the shape of persistently migrating cells in the CPM \cite{Segerer:2015,Thueroff:2017} tends to be fan-like and elongated perpendicular to the direction of motion.
In these previous studies, the elongation increases with the polarizability $\Delta Q \,{=}\, Q - q$ of the cell.
This maximal polarizability translates to a realized polarization strength $p$ of the cells in question.
Therefore, decreasing the polarization strength of the cell will effectively also lead to a rounder cell shape, as happens with decreasing substrate stiffness.
Additionally, one can argue for high substrate viscous friction that the substrate density profile leading to the focussing effect of the cell velocities 'squeezes' the cell together because fewer protrusions occur at the sides of the cell.

\begin{figure}[t]
	\includegraphics{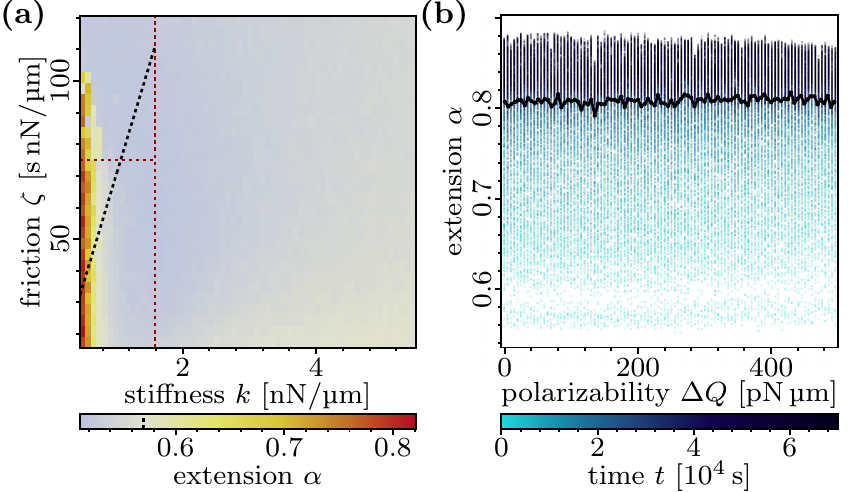}
	\caption{\textbf{Cell extension.}
	\textbf{(a)} Strong cell elongation occurs for low substrate stiffnesses and within a broad range of viscous friction coefficients.
	The red vertical dashed line corresponds to the stiffness $k \,{\approx}\, \SI[per-mode=symbol]{1.58}{\nano\newton\per\micro\meter}$, below which no persistent cell migration occurs.
	The red horizontal dashed line corresponds to the viscous friction $\zeta^\star \,{=}\, \SI[per-mode=symbol]{75}{\second\nano\newton\per\micro\meter}$, above which we on average observe only running states [Fig.~\ref{fig::viscosity_detailed}(d)].
	The black dashed line corresponds to the lower limit of $\zeta^\star$, where we from a simple estimation would expect viscous effects to dominate.
	The color code represents the extension of the cell for a given parameter combination.
	In the color code, the elongation of a cell on a non-deformable substrate is indicated by the dashed line.
	\textbf{(b)} Cell elongation at low substrate stiffnesses does not depend on cell polarizability $\Delta Q \,{=}\, Q-q$.
	The color code represents the current elapsed time of a given data point in the simulation (color bar).
	}
	\label{fig::elongation}
\end{figure}

For low viscous friction and low stiffness of the substrate, we observe a profound elongation of immotile cells.
This observation can be explained in the following way:
Due to the inhibition of the positive feedback mechanism, the cell is unpolarized.
Hence, its behavior is dominated by the substrate density profile alone, and is indeed the same for cells of a wide range of different polarizabilities $\Delta Q \,{=}\, Q - q$ [Fig.~\ref{fig::elongation}], but identical average traction force.
This particular substrate density profile is characterized by an increased substrate density at the short sides of the cell compared to the long sides of the cell.
Hence, the cell has an increased protrusion activity at its short sides and subsequently stretches, as it tries to occupy areas of high substrate density.

\clearpage
\section{Experimental methods}
\label{sec::experimental_methods}

\subsection{Preparation of polyacrylamide substrate}
Acrylamide solutions corresponding to \SIlist{0.2; 1; 2; 3; 7; 15; 34; 100}{\kilo\pascal} polyacrylamide hydrogels were prepared according to previous publications \cite{Tse:2010}.  Briefly, an acrylamide/bis-acrylamide solution (Bio-Rad) was degassed and mixed with 1/100 volume 10\% ammonium persulfate and 1/1000 volume tetramethylethylenediamine (Sigma Aldrich). \SI[per-mode=symbol]{20}{\micro\liter} of solution was pipetted into a single well of an untreated 12 well glass bottom plate (In Vitro Scientific).  A \SI[per-mode=symbol]{12}{\milli\meter} glass coverslip chlorosilanized with dichlorodimethylsilane (Sigma Aldrich) was placed above the acrylamide solution.  Upon polymerization, hydrogels were rinsed and functionalized with photoactivatable Sulfo-SANPAH (Thermo Fisher) before overnight conjugation with \SI[per-mode=symbol]{100}{\micro\gram\per\milli\liter} Collagen type I (Gibco).  Prior to cell culture all hydrogels were UV sterilized.

\subsection{Cell culture}
Human umbilical vein endothelial cells (HUVECs) were cultured in ready-to-use Endothelial Cell Growth Media (PromoCell) with \SI[per-mode=symbol]{100}{units\per\milli\liter} penicillin and \SI[per-mode=symbol]{100}{\micro\gram\per\milli\liter} streptomycin (Gibco).  For live cell imaging, cells were plated at a density of approximately \SI[per-mode=symbol]{2500}{cells\per\centi\meter\square}, or \SI[per-mode=symbol]{10000}{cells\per well}.

\subsection{Microscopy}
Cells on hydrogels were maintained in a microscope-mounted incubator at \SI[per-mode=symbol]{37}{\celsius} and 5\% CO\textsubscript{2}. An AxioVert 200M with Axiovision software (Zeiss) and a PerkinElmer UltraVIEW ERS with Volocity software (PerkinElmer) were used to capture phase contrast images every ten minutes for a period of \SI[per-mode=symbol]{48}{\hour}.

\subsection{Cell shape analysis}
\begin{figure}[t]
	\includegraphics{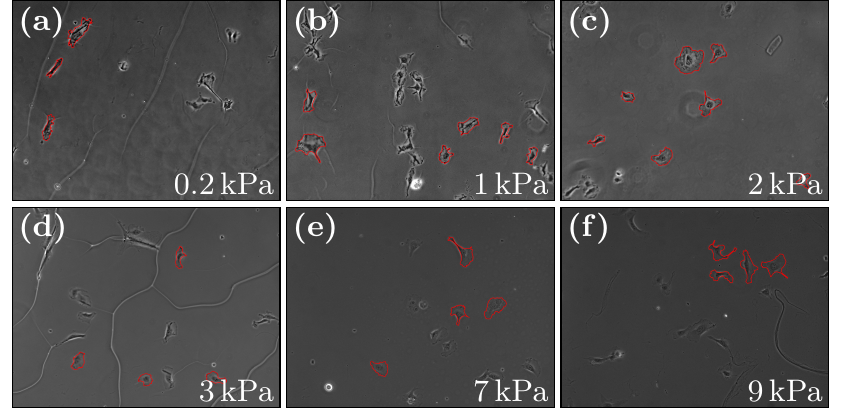}
	\caption{
	\textbf{Manual image analysis procedure.}
	Cells are selected and measured over the course of several frames to average out fluctuations in cell shape.
	Each cell satisfies the following conditions:
	(i) no alignment with the cracks on the substrate;
	(ii) no cell division within this timeframe;
	(iii) distinctness from background and neighboring cells.
	}
	\label{fig::imagej}
\end{figure}

The cell shapes were extracted manually from a subset of the phase contrast images using the software Fiji~\cite{Schindelin:2012ir} [Fig.~\ref{fig::imagej}]. 
Mathematica was used to perform a statistical analysis of the cell shapes~\cite{mathematica}. 
Cells were selected and measured at various times to mitigate fluctuations in cell shape.
Each cell should satisfy three conditions:
\begin{itemize}
	\item no alignment with the cracks on the substrate, as this is a strong bias for the measurement towards elongated cells,
	\item no divisions at the time of measurement, as this is a strong bias for the measurement towards round cells,
	\item good distinguishability of cells from the background and neighboring cells.
\end{itemize}

\subsection{Cell speed analysis}
The cell positions were extracted manually from a subset of the phase contrast images using the software Fiji~\cite{Schindelin:2012ir} and the Plug-In MTrackJ~\cite{Meijering:2012uh} [Fig.~\ref{fig::imagej}].
Mathematica was used to perform a statistical analysis of the cell speeds~\cite{mathematica}.

\clearpage

\end{document}